%% file: main.tex
\documentclass[acmsmall]{acmart}

\usepackage{amsmath}
\usepackage{array}
\usepackage{ulem}

\newcommand\REVA[1]{{\color{black}#1}}
\newcommand\REVB[1]{{\color{black}#1}}

\usepackage{makecell}
\usepackage{multirow}
\usepackage{boldline}
\usepackage{xcolor}
\setcopyright{acmcopyright}
\copyrightyear{2022}
\acmYear{2022}
\acmDOI{XXX}

\acmJournal{TECS}
\acmVolume{123}
\acmNumber{1}
\acmArticle{1}
\acmMonth{1}

\begin{document}

\title{DNN is not all you need: Parallelizing Non-Neural ML Algorithms on Ultra-Low-Power IoT Processors}


\author{Enrico Tabanelli}
\affiliation{%
  \institution{DEI, University of Bologna}
  \streetaddress{viale del Risorgimento 2}
  \city{Bologna}
  \country{Italy}}
\email{enrico.tabanelli3@unibo.it}

\author{Giuseppe Tagliavini}
\affiliation{%
  \institution{DISI, University of Bologna}
  \streetaddress{viale del Risorgimento 2}
  \city{Bologna}
  \country{Italy}}
\email{giuseppe.tagliavini@unibo.it}

\author{Luca Benini}
\affiliation{%
  \institution{DEI, University of Bologna}
  \streetaddress{viale del Risorgimento 2}
  \city{Bologna}
  \country{Italy}}
\email{luca.benini@unibo.it}

\thanks{This work was partially supported by the H2020 "The European PILOT" project (under grant ID 101034126).}

\renewcommand{\shortauthors}{Tabanelli et al.}

\begin{abstract}
Machine Learning (ML) functions are becoming ubiquitous in latency- and privacy-sensitive IoT applications, prompting a shift toward near-sensor processing at the extreme edge and the consequent increasing adoption of Parallel Ultra-Low Power (PULP) IoT processors.
These compute- and memory-constrained parallel architectures need to run efficiently a wide range of algorithms, including key Non-Neural ML kernels that compete favorably with Deep Neural Networks (DNNs) in terms of accuracy under severe resource constraints.
In this paper, we focus on enabling efficient parallel execution of Non-Neural ML algorithms on two RISCV-based PULP platforms, namely GAP8, a commercial chip, and PULP-OPEN, a research platform running on an FPGA emulator.
We optimized the parallel algorithms through a fine-grained analysis and intensive optimization to maximize the speedup, considering two alternative Floating-Point (FP) emulation libraries on GAP8 and the native FPU support on PULP-OPEN.
Experimental results show that a target-optimized emulation library can lead to an average 1.61$\times$ runtime improvement \REVA{and 37\% energy reduction} compared to a standard emulation library, while the native FPU support reaches up to 32.09$\times$ and \REVA{99\%, respectively}.
In terms of parallel speedup, our design improves the sequential execution by 7.04$\times$ on average on the targeted octa-core platforms \REVA{leading to energy and latency decrease up to 87\%}.
Lastly, we present a comparison with the ARM Cortex-M4 microcontroller (MCU), a widely adopted commercial solution for edge deployments, which is 12.87$\times$ slower and \REVA{98\% less energy-efficient} than PULP-OPEN.
\end{abstract}



\maketitle

\input{01-introduction}
\input{02-related_work}
\input{03-background}
\input{04-algorithm_design}
\input{05-experiments}
\input{06-conclusion}

\bibliographystyle{unsrt}
\bibliography{references}

\end{document}

%% file: 01-introduction.tex
\section{Introduction}
%
%
Leading by the recent progress in machine computing power, communication technologies, and big data, Machine Learning (ML) has unveiled cutting-edge breakthroughs in a broad range of domain-specific applications.
As a crucial factor for the widespread use of ML systems, Internet-of-Things (IoT) devices have recently experienced explosive growth, reaching 50B of connected devices in 2020~\cite{evans2011IoT}.
Spanning from Autonomous Driving~ \cite{dogan2011autonomous} to Non-Intrusive Load Monitoring~\cite{tabanelli2021trimming}, ML has become ubiquitous, witnessing a booming of Artificial Intelligence (AI) services and applications~\cite{ml2017ku}.
Due to the proliferation of edge devices, the amount of data generated at the network edge has increased dramatically, reaching 850~ZB of data by 2025~\cite{cisco2011IoT}.
So far, the limited computational capabilities of resource-constrained MCU-based systems have favored offloading data to the cloud for analytics, where computational resources are flexible and virtually unbounded. 
However, the cloud-computing paradigm suffers from scalability issues concerning communication latency, bandwidth, and privacy~\cite{barbera2013offload,sun2014data}.

Latency- (e.g., Autonomous Vehicles) and privacy-sensitive IoT applications (e.g., Health Monitoring Wearable Devices) are prompting a paradigm shift~\cite{sanchez2020tinyml,banbury2020benchmarking,tinyML} toward near-sensor processing at the extreme edge to unleash the potential of ML.
Such applications demand fast and accurate automated decision-making capabilities while handling highly confidential and sensitive customer data. 
Pushing the ML frontiers closer to the information sources promises several benefits, including energy efficiency, data privacy protection, reduced bandwidth costs, and low-latency response~\cite{yu2017survey}.

Unfortunately, moving the intelligence to the edge is non-trivial due to the limited computational capabilities and energy efficiency of resource-constrained IoT devices.
As shown in Table \ref{tab:intro}, modern ML inference tasks run on cloud servers and mobile platforms featuring a peak processing power of up to 38.7~TFLOPS and 155~GFLOPS, respectively.
Instead, the ARM Cortex-M4 MCU represents a widely used platform for edge deployments leveraging a 461000\(\times\) lower computational capability.
Off-the-shelf Deep Neural Networks (DNNs) inference demands hundreds of GFLOPs, largely exceeding typical timing requirements for most applications when executing on state-of-the-art (SoA) single-core MCUs. 
With 3.8~GFLOPS per inference, ResNet~\cite{he2016deep} demands 44.19s running on the ARM Cortex-M4 platform while executing EfficientNet-B0~\cite{tan2019efficientnet} and MobileNet-V2~\cite{sandler2018mobilenetv2} requires 8.45s and 2.33s per inference, respectively.
\begin{table*}[t]
    \centering
    \caption{
    Computational capabilities of ML inference platforms from cloud to edge deployment}
    \renewcommand{\arraystretch}{1.8}
    \resizebox{\textwidth}{!}{
    \begin{tabular}{c c c c c c} 
        \hlineB{3}
        
        & \thead{\textbf{Cloud ML}\\(NVIDIA A100 - Ampere)} & \(\rightarrow\) & \thead{\textbf{Mobile ML}\\(iPhone - Apple A13)} & \(\rightarrow\) & \thead{\textbf{Edge ML}\\(STM32F401 - ARM Cortex-M4)} \\
        
        \hlineB{3}
        
        \textbf{Compute Power} (FLOPS/s) & 38.7T & \(\xrightarrow{\text{250000\(\times\)}}\) & 155G & \(\xrightarrow{\text{1845\(\times\)}}\) & 84M \\
        
        \hlineB{3}
    \end{tabular}
    }
\label{tab:intro}
\end{table*}

Emerging Parallel Ultra-Low-Power (PULP) processors~\cite{gapprocessors,spresensesony} represent an appealing target for TinyML applications since they enable to meet the ML computational constraints in a power envelope of a few milliWatts.
The PULP paradigm builds upon near-threshold computing while leveraging data- and thread-level parallelism to overcome the performance reduction at low operating voltages \cite{mittal2015survey}.
By integrating an I/O-dedicated core with a multi-core Cluster (CL) of processors, this platform offers a flexible software-oriented acceleration for ML and Digital Signal Processing (DSP) tasks.
In this work, we leverage two RISCV-based PULP MCUs to provide proper computing capabilities for ML at the edge.
GAP8~\cite{flamand2018gap8} is a commercial off-the-shelf chip delivering up to 10~GMAC/s (90~MHz, 1.0~V) at the energy efficiency of 600~GMAC/s/W within a worst-case power envelope of 75~milliWatts.
Instead, PULP-OPEN is a research platform running on an FPGA emulator, whose most recent silicon embodiment features a 32.2~GOPS peak performance with a maximum power envelope of 49.4~milliWatts~\cite{rossi2021vega}.

\REVA{
Standard edge-class MCUs usually trade off silicon area and energy efficiency for programmability, limiting the HW resources to the bare minimum to improve the power envelope~\cite{gottscho2017low}. 
At the same time, ML applications demand processing FP workloads since FP support enables satisfying the requirements of dynamic range and precision without intensive numerical tuning.
Due to such tight design and power constraints, small, low-cost IoT cores cannot always afford the cost of a full-fledged HW Floating-Point Unit (FPU).
Several industry-standard STM\footnote{\url{www.st.com/en/microcontrollers-microprocessors/stm32-32-bit-arm-cortex-mcus.html}} and NXP\footnote{\url{www.nxp.com/products/processors-and-microcontrollers/arm-processors:ARM-PROCESSORS}} System-on-Chips (SoCs) integrate FPU-less ARM Cortex-M family cores\footnote{\url{developer.arm.com/Processors/Cortex-M0}} to enable low-power operation. 
Furthermore, commercial devices such as 16-bit PIC and MSP430\footnote{\url{www.ti.com/microcontrollers-mcus-processors/microcontrollers/msp430-microcontrollers/products.html}} MCUs, along with Xtensa L106 core embedded into ESP8266 SoCs\footnote{\url{www.espressif.com/en/products/socs/esp8266}}, follow this trend.
These FPU-less devices implement FP computation with SW FP emulation.
Deriving the fixed-point variant of FP algorithms is highly time-consuming~\cite{doris2013profile} and requires additional analysis that takes up 30\% of the overall development time~\cite{menard2006floating}. 
In addition, fixed-point computations are deeply susceptible to quantization effects, thus making FP conversion error-prone and challenging~\cite{christensen2006fixed,menard2008accuracy,chang2008fft}. 
Edge applications constrained by tight resource budgets and short time-to-market would be negatively impacted by adopting fixed-point arithmetic. 
In this scenario, using fast FP SW emulation libraries brings several benefits by decreasing development time and enabling fast time-to-market. 
Parallelizing emulated FP workloads on multi-core ULP devices can dramatically reduce the runtime overhead introduced by FP SW emulation while still meeting the power budget of TinyML applications. 
In this paper, we consider two alternative FP emulation libraries on GAP8 \REVB{since this target} does not offer FPU-native support. 
libgcc provides a set of standard low-level routines to handle arithmetic operations not natively supported by the target platform. 
We also deploy RVfplib, which consists of a library optimized for FP arithmetic emulation on 32-bit RISCV processors~\cite{perotti2021rvfplib}.}

In recent years, academic and industrial researchers have focused their interest on DNNs, introducing novel topologies to improve accuracy and efficiency, customizing hardware designs and instruction set architectures (ISA) to DNN execution~\cite{capra2020hwswnn}.
At the same time, Non-Neural ML kernels have been partially neglected by the TinyML research community. 
Nevertheless, for a wide range of applications, these algorithms lead to an accuracy comparable with SoA DNNs while demanding lower computing capabilities.
Greeshma et al. \cite{greeshma2019fashion} achieve near-SoA accuracies on the \REVA{Fashion-}MNIST dataset~\cite{lecun1998gradient} deploying a set of Non-Neural ML algorithms: linear Support Vector Machine (SVM) and Random Forest (RF) attain up to 97.3\% accuracy. At the same time, Logistic Regression (LR) and k Nearest-Neighbor (kNN) reach 91.7\% and 95.9\%, respectively.
Thus, Non-Neural ML algorithms represent an important target for optimized deployment on PULP-class devices for TinyML.
In this scenario, the primary goal of our work is to optimize the parallel design of a set of Non-Neural MLoptimize the parallel design of a set of Non-Neural ML algorithms to run efficiently on two RISCV-based PULP MCUs.

The main contributions of this paper are:
\begin{itemize} 
    \item \REVA{We optimize the sequential and parallel design of six widely utilized Non-Neural ML algorithms, maximizing the Cycles per Instructions (CPI) metric on two RISCV-based PULP MCUs. We provide a detailed experimental assessment that explains the architectural factors limiting the performances at the core- and system-level. We compute the floating-point operations (FLOP) intensity for each kernel to describe in-depth the achieved performance with alternative FP emulation supports and FPU-native system. We also report the theoretical speedup following Amdahl’s law to motivate the structural limitations on parallel performance.}
 	\item \REVA{We compare the kernel execution time when running on a single-core configuration, leveraging alternative floating-point (FP) emulation libraries on GAP8 and the FPU-native support on PULP-OPEN. We also report code size, energy consumption, and latency for each algorithm and platform configuration. The experimental evaluation shows that the target-optimized RVfplib library achieves an average 1.61\(\times\) speedup and 6.24\% code size reduction compared to the standard libgcc emulation support. Adopting the fast SW emulation library also enables a 37\% energy reduction. The FPU-native support reaches up to 32.09\(\times\) speedup and 41.71\% code size decrease compared to libgcc emulation.}
	\item \REVA{We examine the 1-vs-8 cores parallel speedup achieved on the targeted PULP platforms, considering FP emulation on GAP8 and FPU-native support on PULP-OPEN. The results reveal that our optimized parallel design allows achieving near-ideal speedups for Non-Neural ML kernels, ranging from 6.56\(\times\) to 7.64\(\times\) compared to a single-core execution. We also report an energy and latency reduction of up to 87\%.}
	\item \REVA{We compare the Non-Neural ML algorithms execution time running on PULP-OPEN and the ARM Cortex-M4 MCU. The experimental results demonstrate that a single-core PULP-OPEN configuration leads to speedups ranging from 1.36\(\times\) to 2.39\(\times\) compared to Cortex-M4 deployment, along with 85\%-89\% average energy and latency reductions. While fully leveraging the PULP-OPEN 8-core CL diminishes the computing time by more than one order of magnitude, between 9.27\(\times\) and 15.85\(\times\). We also provide parallel design energy and latency improvements, which reach up to 98\% decreases compared to Cortex-M4. }
	
\end{itemize}

%% file: 02-related_work.tex
\section{Related work}
\REVA{
\subsection{NN Tools And Libraries}
The current generation of SW frameworks and tools for TinyML mainly focuses on neural ML algorithms deployment on SoA single-core MCUs.
A significant representative of this trend is CMSIS-NN \cite{lai2018cmsis}, a software library including a set of kernels developed to maximize the performance and minimize the memory footprint of NNs on ARM Cortex-M family cores.
X-CUBE-AI \cite{x-cube-ai} from STMicroelectronics\footnote{\url{https://www.st.com/content/st_com/en.html}} converts pre-trained NNs exported from common DL frameworks into a pre-compiled library optimized on computation and memory targeting STM32 MCUs.
By addressing optimal memory tiling and efficient data transfers, the AutoTiler tool from GreenWaves Technologies\footnote{\url{https://greenwaves-technologies.com/}} generates code from pre-trained DNNs supporting the execution on the RISCV-based multi-core MCU GAP8.}

\subsection{Non-Neural ML Libraries}
While the aforementioned solutions enable deploying NN workloads on several MCUs, they do not support generating code for pre-trained Non-Neural ML algorithms.
Consequently, several works have been proposed recently from the industry and open-source domain to support Non-Neural kernels inference at the edge.
CMSIS-DSP is a software library including a comprehensive set of DSP functions optimized by ARM for various Cortex-M processors with FP support. 
Recent versions of CMSIS-DSP add new functions support for Non-Neural ML algorithms, including alternative SVM kernels, a Naive Bayes estimator, and distance functions for clustering algorithms.
The TinyML paradigm includes a set of techniques to integrate ML algorithms within resource-constrained MCUs \cite{sanchez2020tinyml}.
Yazici et al. \cite{yazici2018edge} implement SVM and RF models on a Raspberry Pi platform, reporting accuracy between 82\% and 96\% and an execution time of around 5 seconds to perform inference on 100 instances.
However, the Raspberry Pi platform has a power envelope of 2-5 Watts~\cite{bekaroo2016pi}, which far exceeds the few milliWatts power budget of TinyML applications.
Furthermore, \cite{yazici2018edge} does not provide any insight into the algorithm design.
Edge Machine Learning (ELM) \cite{sakr2020machine} consists of an open-source ML framework targeting STM32 edge devices, implementing linear kernel SVM, RF, Decision Tree (DT), and k-NN.
Instead, \textit{MicroML} \cite{microml} and \textit{emlearn} \cite{emlearn} are Python modules that extend the Scikit-learn library to generate Non-Neural ML algorithms targeting edge MCUs, including SVM, RF, DT, and naïve Gaussian Bayes algorithms. 
These libraries provide platform-independent C implementations for a wide range of target MCUs, without dependencies with external libraries and with integer/FP arithmetic support.
However, these solutions do not provide platform-specific optimizations necessary to achieve peak performance at the edge and do not support parallel execution on multi-core Ultra-Low-Power (ULP) processors.

\REVA{
\subsection{Non-Neural ML Parallelization}
In the last years, several works have been proposed to tackle the efficient parallelization of Non-Neural ML algorithms on many- and multi-core architectures~\cite{almansoor2020parallel,senagi2022parallel,liu2019parallel}.
However, such approaches target high-end platforms leveraging resources unavailable on MCU-class devices and fail to meet the limited TinyML budget.
They also primarily focus on accelerating the algorithms training phase by deploying multi-level parallelism with complex memory hierarchies provided by these architectures. 
In~\cite{you2014mic}, the authors designed a highly efficient parallel SVM training on x86-based many-core architectures, achieving up to 84$\times$ and 47$\times$ speedups w.r.t. LIBSVM on the Intel Xeon Phi co-processor and Ivy Bridge CPU. 
Unfortunately, the design utilizes task- and data-level parallelism by leveraging multiple threads and a Vector Processing Unit (VPU) to reach satisfactory performances. 
Parallel Ultra-Low-Power platforms usually limit the HW resources to meet a power envelope of a few milliWatts, thus not supporting standard Multi-Threading programming models and large vector units.
Zhu et al.~\cite{zhu2019high} compared an OpenMP- and OpenCL-based parallel learning to Rank SVM for multi-core CPUs and GPUs, proving that OpenCL reaches 7.8$\times$ and 19.3$\times$ speedup on such platforms.
However, OpenCL parallel programming model leverages features not supported by MCU-class devices, such as shared virtual memory and dynamic parallelism.
By conducting a comprehensive study of parallel LR training, Ma et al.~\cite{ma2019stochastic} 
reduced the computing time by 200$\times$ and 500$\times$ on an Intel multi-core CPU and NVIDIA GPU.
The approach relies on techniques generally not supported by our edge devices, such as multi-threading, load balancing to allocate virtual threads, and minimization of thread creation/destruction events.
}

\subsection{HW/SW Optimizations}
In the last decade, researchers have proposed specialized designs to reduce the inference costs of ML algorithms.
Microsoft released the EdgeML\footnote{\url{https://github.com/microsoft/EdgeML}} library, which consists of novel Non-Neural ML algorithms suitable for severely resource-constrained edge and IoT devices.
For example, ProtoNN \cite{gupta2017protonn} is a kNN-based algorithm designed to reduce model size and execution time on IoT devices with less than 32~kB memory and a frequency of 16~MHz.
While ProtoNN efficiently handles extensive datasets obtaining SoA accuracy, its related optimization problem is non-convex, requiring the adoption of stochastic gradient descent (SGD) with iterative hard thresholding to perform training.
Bonsai \cite{kumar2017resource} is a tree-based algorithm designed to guarantee efficient prediction on IoT devices such as the Arduino Uno board, operating at 16~MHz with no FPU-native support, 2~KB RAM, and 32~KB read-only flash.
Bonsai learns a single, shallow, sparse tree in which both internal and leaf nodes make non-linear predictions: the overall prediction is computed as the sum of the individual predictions along the path traversed by an input sample. 
This approach reduces the model size compared to the solution that employs independent classifiers in the leaf nodes. 
%
%
Since MCU-based devices for IoT applications often do not integrate an FPU, Gopinath et al. \cite{gopinath2019kbsized} proposed a framework that generates efficient fixed-point code for ML inference at the edge.
Moreover, this approach requires expressing the ML algorithm in a domain-specific language and using a custom compiler.
\REVB{
Mahajan et al.~\cite{mahajan2016tabla} describe a template-based framework to accelerate a set of learning algorithms (including LR and SVM) on FPGA.
FPGA acceleration is a viable approach in many domains, but its power budget is too high for ULP processing at the edge of the IoT.
}

In this paper, we optimize the parallel design of six very common Non-Neural ML kernels \cite{MAHDAVINEJAD2018161,merenda2020edge} achieving peak performance on two RISCV-based multi-core PULP MCUs. 
We designed the algorithms using the C programming language standard while integrating low-level platform-dependent optimizations into the runtime.
Following, we deeply detail the design through a fine-grained analysis describing the parallelization patterns and memory access optimizations adopted.

%% file: 03-background.tex
\section{Background}
\label{Background}
This Section briefly describes the target MCUs and the software ecosystem deployed in this work, along with a motivations discussion presented in Section~\ref{Motivations}. 
The PULP platform will be presented in Section~\ref{PULP_Platform}, while GAP8 and PULP-OPEN in Sections~\ref{GAP8} and \ref{PULP_OPEN}, respectively.
Along with this, we report in Section~\ref{FP_Emulation_Libraries} the two FP emulation libraries deployed to enable FP computations on architectures with no FPU-native support.
Finally, in Section~\ref{Programming_and_Compilation_Toolchain}, we introduce the software stack and parallel programming model used to achieve fine-grained data- and thread-level parallelism.
\subsection{Motivations}
\label{Motivations}
\REVB{
SoA DNNs achieve the highest accuracy in many application fields, including Keyword Spotting, Computer Vision, and Anomaly Detection.
However, their higher performance comes with a price of computational complexity, hampering their applications in many resource-constrained platforms, such as MCU-based IoT devices.
Moreover, DNN performs only marginally better than tree-based models in some application fields (e.g., energy prediction models \cite{ahmad2017trees}).
For these reasons, non-neural ML techniques remain widely used for ultra-low-power and tightly resource-constrained near-sensor processing applications. In fact, a few commercial smart sensors, such as the LSM6DSOX system-in-package by STMicroelectronics, feature an embedded hardware processing engine accelerating DTs for "in-sensor" processing and classification.
}

\REVB{To quantitatively assess the complexity vs. accuracy tradeoff on open benchmarks, we analyzed the accuracy achieved by Non-Neural ML algorithms and SoA DNNs while comparing the computational complexity at inference time in terms of Multiply-and-Accumulate (MAC) operations.}
The study has been conducted on three widespread industrial and commercial use cases: Keyword Spotting, Image Classification, and Anomaly Detection. 
Using the well-known MLPerf Tiny benchmark suite~\cite{banbury2021mlperf}, we considered Speech Commands, CIFAR-10, and ToyADMOS datasets, and DS-CNN, ResNet-8, and FC-Autoencoder (FC-AE) as SoA DNN references.

\begin{figure}[htp!]
    \centering
    \captionsetup{justification=centering}
    \includegraphics[width=0.9\linewidth]{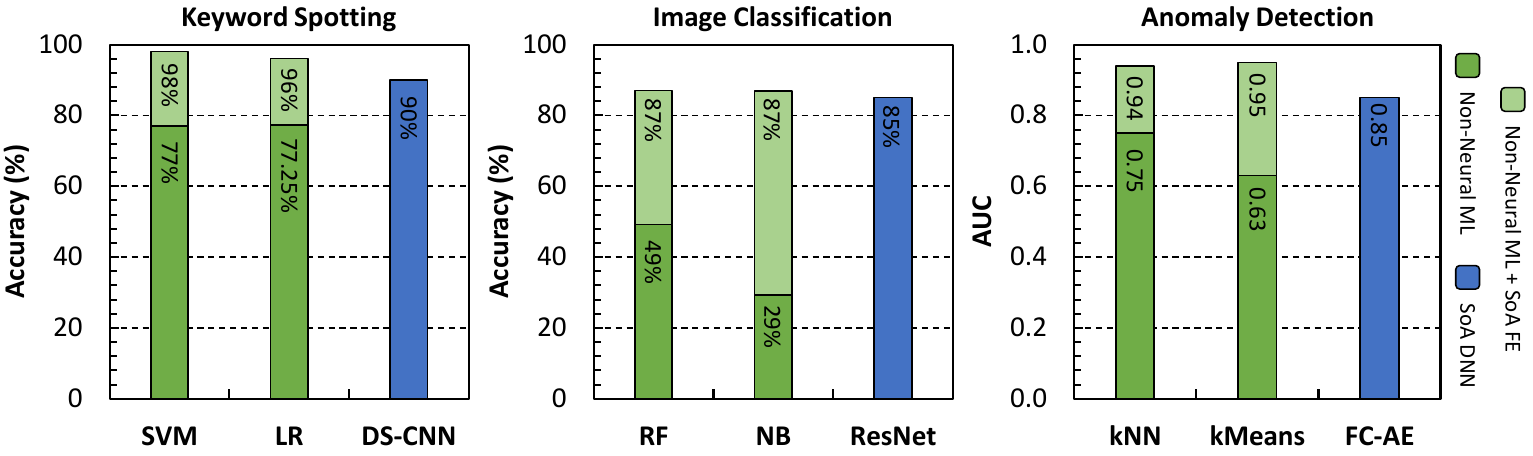}
    \vspace{-0.2cm}
    \caption{\REVA{Non-Neural ML vs SoA DNNs} \REVB{Top-1} \REVA{accuracy. Abbreviations: Feature Extractor (FE).}}
    \vspace{-0.2cm}
    \label{fig:DNNs_vs_NNML_ACC}
\end{figure}
\begin{figure}[htp!]
    \centering
    \captionsetup{justification=centering}
    \includegraphics[width=0.9\linewidth]{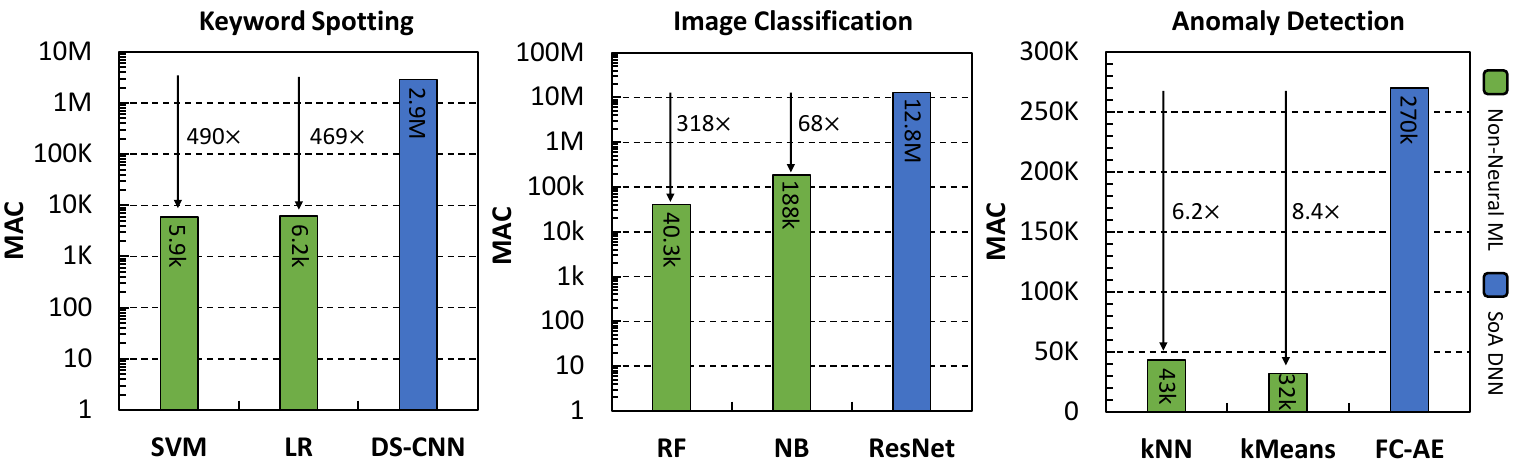}
    \vspace{-0.2cm}
    \caption{\REVA{Non-Neural ML vs SoA DNNs computational complexity.}}
    \vspace{-0.2cm}
    \label{fig:DNNs_vs_NNML_MACs}
\end{figure}

\REVA{As shown in Figure~\ref{fig:DNNs_vs_NNML_ACC}, we executed GEMM-based Non-Neural ML algorithms on the Speech Commands dataset for the Keyword Spotting task. 
The DS-CNN architecture reaches 90\% accuracy but at a higher cost of 2.9~MMACs per inference, as depicted in Figure~\ref{fig:DNNs_vs_NNML_MACs}.
Leveraging Non-Neural ML models enables lowering the computational complexity to only 6~kMACs with a 490$\times$ speedup, still reaching \REVB{an acceptable} 77\% accuracy.
It is important to notice that the accuracy of DNNs on these tasks keeps increasing, but at the same time non-neural ML approaches are also getting better.
}
\REVB{
In recent years, academic researchers have also focused on leveraging custom feature extractors on top of SVM and LR. 
On the Speech Commands dataset, Huh et al.~\cite{huh2021metric} reached 98\% accuracy by changing the loss functions from the classification loss to a range of metric learning objectives and then training a one-vs-one SVM kernel.
On the NOSS benchmark suite, Shor et al.~\cite{shor2022universal} trained an LR classifier on time-averaged representations achieving 96\%.
}

\REVA{
To assess Non-Neural ML algorithms performance in image classification, we trained RF and NB models on CIFAR-10 achieving up to 50\% accuracy, while ResNet-8 architecture leads to 85\%. 
However, adopting Non-Neural ML kernels decreases the computational complexity by up to 318x, requiring only 40.3~kMACs per inference against the 12.8~MMAC demanded by ResNet-8.
Furthermore, many works have investigated the use of CNN-based feature extractors to pre-process image pixels leading to astonishing performances when coupled with Non-Neural ML kernels. 
}
\REVB{
Liu et al.~\cite{liu2019fusing} reached 87.2\% accuracy on CIFAR-10 training a set of DTs with the feature extracted from the last fully-connected layer of a ResNet; using NB, they achieved 86.6\% accuracy.
}

\REVA{
Lastly, we evaluated performances in the Anomaly Detection scenario by comparing kNN and kMeans kernels against the FC-Autoencoder architecture on the ToyADMOS dataset.
The SoA DNN achieves a 0.85~AUC score requiring 270~kMACs to detect abnormal input data. 
At the same time, Non-Neural ML algorithms reduce computing time by 6.2x with merely 43~kMACs per inference and still lead to \REVB{an acceptable} 0.75~AUC. 
Several works also studied alternative feature extractors to improve the performance of Non-Neural ML kernels in Anomaly Detection. 
Durkota et al.~\cite{durkota2020neuron} reach up to 0.94~AUC by deploying a Siamese Network to extract features on top of the kNN model while using the Mutual Information technique enables reaching 0.95~AUC with k-Means~\cite{zhao2021spatiotemporal}. 
}

\REVB{
To summarize the discussion, SoA works on alternative feature extractors have proved that Non-Neural ML algorithms can still compete with SoA DNNs in terms of accuracy in several industrial scenarios, often achieving significant reductions in computational and memory footprints.
Since low-cost IoT devices are subject to tight memory and compute constraints, the efficient acceleration of these kernels is practically a relevant target and will remain so in the near future.
}
This paper focuses on enabling efficient parallel execution of Non-Neural ML algorithms on two RISCV-based PULP platforms.
\subsection{PULP Platform}
\label{PULP_Platform}
PULP is a RISCV-based open-source platform\footnote{\url{https://github.com/pulp-platform}} built on the near-threshold computing paradigm~\cite{mittal2015survey}.
The ultra-low-power design allows outstanding energy efficiency while data- and thread-level parallelism overcome the performance reduction at low operating voltages.

\begin{figure}[t]
    \centering
    \captionsetup{justification=centering}
    \includegraphics[width=0.9\linewidth]{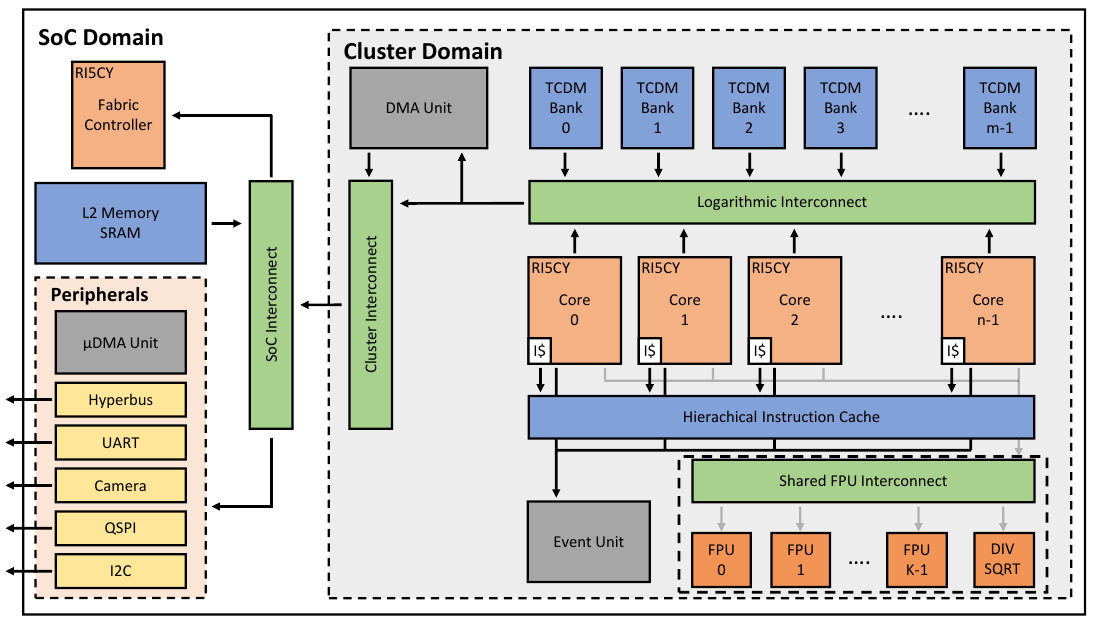}
    \caption{Top-level view of the PULP platform System-on-Chip.}
    \label{fig:PULP_Platform}
\end{figure}

Figure~\ref{fig:PULP_Platform} depicts the PULP System-on-Chip (SoC) top-level design. The microarchitecture is divided into two isolated voltage and frequency domains, managed by DC/DC and Frequency-Locked Loops (FLLs): the Fabric Controller (FC) and the Cluster (CL).
The PULP CL consists of a configurable number of RI5CY cores, a RISCV-based processor featuring a 4-stage in-order single-issue pipeline, and supporting the RV32IMCXpulpV2 Instruction Set Architecture (ISA).
The standard RV32IMC ISA provides support for integer, compressed, and multiply/divide instructions.
Instead, the XpulpV2 extension enables highly energy-efficient computations with custom ML- and DSP-centric instructions.
For that purpose, XpulpV2 includes hardware loops, post-incrementing load/store, multiply-add instructions, fixed-point, bit-manipulation, and single instruction multiple data (SIMD) support down to 8bit packed data.

The PULP CL replaces traditional data caches with a Tightly Coupled Data Memory (TCDM) to reduce energy and area consumption while leveraging DSP data access pattern predictability.
The memory acts as a size-configurable multi-banked scratchpad memory (SPM) with a banking factor of two (i.e., 8 banks for the 4-cores configuration), enabling shared-memory parallel programming models such as OpenMP~\cite{dagun1998openmp}.
A single-cycle latency word-level interleaved logarithmic interconnect allows data sharing between TCDM and cores with a low average contention rate.
The CL features a hierarchical instruction cache (I\$) consisting of a first private level and a second shared one.
This design provides optimal performances and energy efficiency in fetching data-parallel code, reducing instruction misses, and leveraging the SIMD nature of most near-sensor processing applications.

A custom Hardware Synchronization Unit (Event Unit) implements low-overhead support for fine-grained parallelism, providing fast event management, parallel thread dispatching, and synchronization.
The Event Unit also provides high-energy efficiency by utilizing power-saving policies when cores are in the idle state.
The cores waiting for a synchronization barrier or an event are taken to a fully clock-gated state, thus zeroing the dynamic energy consumption.

On the SoC level, PULP features a RI5CY core and a multi-channel I/O \(\mu\)DMA to manage data transfers and minimize the core workload when performing I/O.
A 15-cycle latency multi-banked SPM memory acts as an L2 hierarchy level that serves the CL data bus, the I\$ refills, and the CL DMA unit. 
The SoC also features a comprehensive set of peripherals enabling parallel capture of images, sounds, and vibrations, for use in smart applications such as speech recognition and object detection.

\subsection{GAP8}
\label{GAP8}
GAP8~\cite{flamand2018gap8} is a commercial SoC for IoT applications, embedding a RISC-V multi-core processor derived from the PULP open-source computing platform.
The SoC leverages a single-core FC coupled with an octa-core CL, enabling AI workload at the edge.

The single-core system acts as an advanced MCU in charge of controlling all the SoC operations while fetching instructions from a 4~kBytes I\$.
Featuring a 512~kB L2 memory reachable by each core and a private 16~kB L1 memory, the FC domain includes a ROM memory to store the primary boot code.
An 800~Mbit/s Double-Data Rate (DDR) Hyperbus interface enables extending the on-chip memory, while a multi-channel \(\mu\)DMA permits hiding L3 data transfer cost.
A set of peripherals (i.e., QuadSPI, I2C, 4I2S, CAM, UART, PWM, GPIOs, JTAG) enables the acquisition of several signals featuring high bandwidth and efficiency.

On the CL side, the SoC integrates 8 identical RI5CY cores with a 16~kB 2-level shared I\$ and a 64~kB multi-banked TCDM.
Offloading highly compute-intensive kernels allows up to 10~GMAC/s (90~MHz, 1.0~V) at the energy efficiency of 600~GMAC/s/W within a worst-case power envelope of 75~mW.
Furthermore, the extremely energy-efficient design enables 3.6~$\mu W$ power consumption when in deep-sleep mode.

\subsection{PULP-OPEN}
\label{PULP_OPEN}
PULP-OPEN is a research-oriented platform based on the PULP project, tailored for applications in the domain of near-sensors computing. 
The platform reflects the GAP8 architecture and microarchitecture, with the addition of FPU native support.

The PULP-OPEN CL integrates FPnew~\cite{mach2021fpnew}, a parametric open-source FPU leveraging the insertion of any number of pipeline stages and supporting a wide variety of standard and custom FP formats.
In this work, we deploy four FPnew instances shared among the eight cores of the CL, each presenting one pipeline stage.
The shared FPU provides support for IEEE 754 single- (FP32) and half-precision floats (FP16), along with custom 16-bit bfloats (FP16alt).
Moreover, the architecture implements SIMD vectorization, vectorial conversions, and data packing/unpacking.

\begin{figure}[t]
    \centering
    \captionsetup{justification=centering}
    \includegraphics[width=0.8\linewidth]{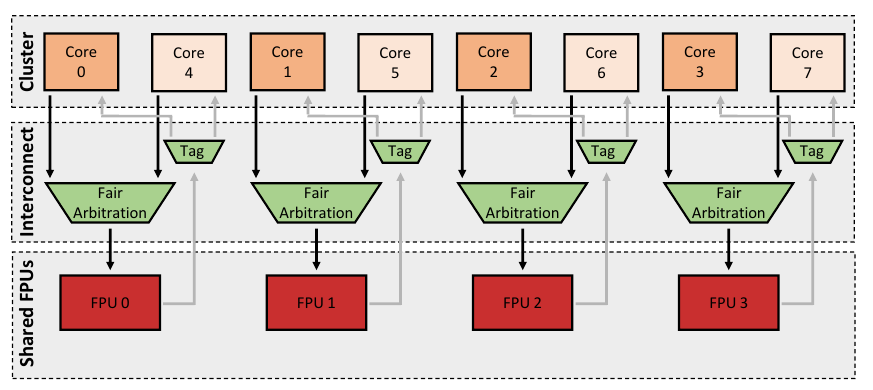}
    \caption{Top-level design of the PULP FPU sub-system}
    \label{fig:FPU}
\end{figure}
Figure~\ref{fig:FPU} depicts the top-level design of the shared FPU exploited in this work.
A logarithmic tree interconnect links individual FPU instances with two cores, enabling sharing FPUs among different cores with total transparency at the software level.
The static mapping of FPUs allows cores to always access the same physical FPU instance.
At the core side, the interconnect interface overrides the FPU during the execution stage, simulating a core-private block. 
An Auxiliary Processing Unit (APU) interface connects the FPU instances to the cores, leveraging ready/valid protocol with a round-robin policy and communicating with the processor execute pipeline stage.
In the case of simultaneous access to the FPU, the system propagates the ready signals to only one processor and stalls the pipeline of the competing core. 
The FPU utilizes a connection scheme with interleaved allocation to decrease access contentions in unbalanced workloads.

\subsection{FP Emulation Libraries}
\label{FP_Emulation_Libraries}
In this work, we deploy FP32 as the standard data format for computations. 
To enable the execution of FP32-based algorithms on GAP8, we perform FP computations employing a standard and a custom FP emulation library.

The GNU Compiler Collection (GCC) provides a low-level runtime library called libgcc. 
The routines integrated into the library handle arithmetic operations not natively supported by the target processor. 
The GCC compiler automatically creates calls to libgcc routines or inlines the code when the target benchmark includes operations with no HW-native support.
In particular, libgcc includes a set of FP IEEE-754 compliant routines supporting single- and double-precision data formats, with a wide variety of arithmetic, conversion, comparison, and advanced software-emulated operations.

To reduce the overhead when executing FP-based kernels on GAP8, we also use RVfplib~\cite{perotti2021rvfplib}, a custom RISCV-based IEEE-754 compliant library optimized for FP arithmetic on 32-bit integer processors. 
The library provides two versions targetting code size and performance optimization compatible with RV32IMC processors. 
In this work, we use the RVfplib version optimized for faster code execution. 
With the support for standard FP32 and FP64 data formats, RVfplib provides target-optimized software routines for conversion, arithmetic, and comparison operations.

\subsection{Programming Model and Compilation Toolchain}
\label{Programming_and_Compilation_Toolchain}
An efficient and low-overhead software stack is mandatory to fully leverage the CL compute power. In this work, we use the PULP open-source software ecosystem\footnote{\url{https://github.com/pulp-platform/pulp-sdk}}, which provides a parallel programming model and compiler support for both targets.

The PULP toolchain provides compiler support for GAP8 and PULP-OPEN platforms. 
It includes an extended version of GCC 7.1 supporting the XpulpV2 extension along with a set of custom relocation schemes supported by the linker.
After loading the code program into L2 memory, the FC executes the application from the entry point and offloads compute-intensive kernels to the CL. 


A Hardware Abstraction Layer (HAL) provides access to low-level resources to explicit the parallel computing paradigm. 
The core identifier allows scheduling the parallel workload among the workers leveraging data- and thread-level parallelism.
An inter-core synchronization is mandatory to ensure correct results in the shared-memory programming model. 
Thus, the CL architecture provides specialized HW support for optimized synchronization primitives, such as barriers and critical sections, to orchestrate the execution flow. 
The OpenMP programming model is also available but implies higher overhead costs than HAL primitives. 
In this work, we focused on maximizing Non-Neural ML algorithms execution performance; hence, we used the lower-level HAL for our experimental assessment.




%% file: 04-algorithm_design.tex
\newcommand{\R}{\mathbb{R}}

\section{Algorithm Design}
\label{Algorithm_Design}
%
In this section, we present the design of six key Non-Neural ML algorithms optimized for parallel execution on the two RISCV-based PULP platforms. After giving an introductory description of the mathematical fundamentals, we thoroughly detail the parallelization strategy used to dispatch the CL workload efficiently. 
We also report the fine-grained analysis and intensive optimization to maximize the speedup.
For simplicity, we grouped the algorithms based on their mathematical formulation and parallelization nature:
\begin{itemize}
    \item General Matrix Multiply based (GEMM-based): LR and SVM.
    \item Gaussian Naive Bayes (GNB). 
    \item Metric Space based (MS-based): kNN and K-Means.
    \item Independent Tasks based (IT-based): RF.
\end{itemize}

To break the TinyML memory bottleneck on resource-constrained devices, the research community usually leverages novel techniques such as optimal double-buffering and memory tiling~\cite{nntile2020zhuo,burrello2021dory}.
We optimized the algorithms as stand-alone kernels fine-grained tuned to process in parallel data placed in L1 memory.
An external double-buffering wrapper enables using L2 memory when data do not fit L1, overlapping L1-L2 memory transfer operations, and kernel processing with almost zero cycles overhead. 
Lastly, we find an optimal tiling strategy for each algorithm fine-tuning the memory accesses to maximize data reuse and performance.

\begin{figure}[t]
    \centering
    \captionsetup{justification=centering}
    \includegraphics[width=0.7\linewidth]{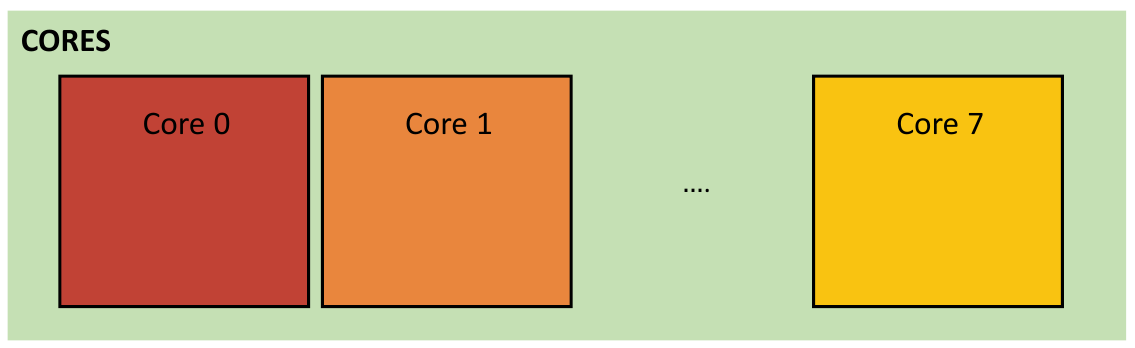}
    \caption{Cores coloring used to mark related processing data.}
    \label{fig:CL_CORES}
\end{figure}

In this section, we detail the design of the stand-alone kernels optimized to run efficiently in parallel onto the octa-core CL.
\REVA{The colors used in the following figures depend on the data associated with each core, as depicted in Figure~\ref{fig:CL_CORES}. We use a specific color for the memory data read by a particular core. Since sequential operations imply executing with a single core, we arbitrarily selected core 0 to execute sequential operations and colored the read memory data in red.}
For each algorithm, we consider a training dataset \(A\) consisting of \(N_{train}\) \(d\)-dimensional samples and \(N_{class}\) classes.
To describe the parallelization schemes, we utilize bold capital and lowercase letters to represent matrices and vectors, while lowercase symbols depict scalar variables. 
%
\REVA{
\subsection{Parallelization Approach}
\label{par_approach}
The OpenMP~\cite{chandra2001parallel, tagliavini2018unleashing} paradigm is a widely-adopted parallel programming model for shared-memory multi-core MCU platforms, and it has already been demonstrated in the context of embedded systems~\cite{munera2020towards,chapman2009implementing,agathos2013deploying} and TinyML applications~\cite{patel2015survey,padilla2020detection,huang2012parallelizing}. 
However, this programming model leads to unavoidable overheads in distributing the workload and orchestrating communication/synchronization among the workers~\cite{furlinger2006analyzing}. 
Minimizing such runtime overheads is crucial to enabling fine-grained parallelism on ULP multi-core platforms. 
Furthermore, TinyML applications have small workloads implying relatively short parallel regions (just a few tens of cycles), making it challenging to amortize overheads. 
The SPMD parallel paradigm~\cite{darema2001spmd} is an alternative approach requiring more programmer effort than OpenMP since it requires modifying the source code and dealing with low-level details (e.g., inter-core synchronization, critical sections, and shared/private variables allocation). 
Nevertheless, the SPMD paradigm enables fine-grained parallelism due to a higher runtime control, leading to less overhead than a traditional OpenMP. 
Montagna et al.~\cite{montagna2021streamlining} compared the two paradigms and proved that a bare-metal SPMD runtime achieves a 178\% runtime improvement compared to a baseline OpenMP on multi-core ULP MCUs. 
Based on this evidence, our work focuses on providing an optimized SPMD version of the code.
}

\REVA{
To further improve the parallel runtime approaching ideal performances, we leverage HW-specific optimizations for core idling and synchronization. 
GAP8 and PULP-OPEN Clusters integrate a multi-core Event Unit (EU) optimized to accelerate key data-parallel patterns execution, such as barriers and locks, while supporting power-saving policies to put cores in idle state. 
The EU is a lightweight HW block designed to enable fine-grained parallelism that aims to achieve minimum synchronization overhead in terms of cycles and energy. 
Due to its efficient HW design, executing barriers and critical sections with the 8-core Cluster configuration requires 6 and 50 Cycles, respectively. 
The barrier and mutex extensions correspond to the parallel and critical section constructs fundamental in most parallel programming models. 
Thus, leveraging EU HW-specialized support is key to drastically reducing the synchronization overhead in parallel programming primitives. 
In our work, we access low-level resources leveraging a Hardware Abstraction Layer (HAL).
}

\subsection{Horizontal and Vertical Workload Distribution}
\label{hor_vs_ver_par}
We introduce two data partitioning schemes adopted in the rest of this section to achieve optimal performance on multi-core platforms, namely horizontal and vertical workload distribution.

As a common pattern, ML workloads include an operation between a \(r\,\times\,c\) matrix \(M\) and a \(c\)-dimensional input vector \(x\), leading to a scalar value \(y\).
In this scenario, programs can conveniently exploit data-level parallelism: a workload distribution strategy splits data into chunks, and each core executes the same code on a different chunk.
This method has an associated overhead since it implies the computation of core-dependent loop bounds. Since this overhead is constant, its impact decreases as the chunk size increases.

Depending on \(r\) and \(c\) dimensions, selecting a partitioning strategy mapped onto horizontal or vertical stripes of the matrix operand could significantly improve CL utilization.
Having \(r >> c\) favours a \REVA{vertical} decomposition.
The strategy involves partitioning \(r\) rows into \(n_{cores}\) chunks consisting of \(r/n_{cores}\) elements.
Instead, \(c >> r\) promotes a \REVA{horizontal} decomposition.
Following the approach, each core computes on \(r\) vectors of dimension \(c/n_{cores}\).

\subsection{GEMM-based Algorithms}
\label{General_Matrix_Multiply_based_Algorithms}
Below, we describe the algorithms based on the GEMM function, a Basic Linear Algebra Subprograms (BLAS) routine largely deployed in statistics and ML. 
As reported in Eq.~(\ref{eq:2}), GEMM-based algorithms leverage the product between two input matrices \(A\) and \(B\), while \(C\) represents a pre-existing matrix overwritten by the output.
\begin{equation}
    C^{m\,\times\,n} = \alpha \cdot A^{m\,\times\,k} \times B^{k\,\times\,n} + \beta \cdot C^{m\,\times\,n}\label{eq:2}
\end{equation}
\(\alpha\) and \(\beta\) are scalar inputs that enable the plain product \(A \times B\) and the output matrix \(C\) accumulation.

LR and SVM present an analogous inference scheme consisting of a GEMM computation performed between the input vector \(x\) and the matrix \(W\) while alternative activation functions process the output.

\subsubsection{Logistic Regression (LR)}
\label{LR}
LR is a supervised ML algorithm for binary classification, which leverages a logistic function to model output probabilities \cite{cramer2002lr}.
While Linear Regression applies an interpolation between points by avoiding distinguishing classes, LR deploys the logistic function to squeeze the linear output between 0 and 1, thus returning the class probability.
Due to its high classification performance and straightforward interpretability, the model has been widely adopted across several real-world scenarios, such as intrusion detection \cite{ioannou2018wsn} and anomaly detection \cite{HASAN2019100059}.

As reported in Eq.~(\ref{eq:lr_decision_function}), LR binary decision function leverages the weighted sum between \(x\) and the real-valued \(d\)-dimensional weights vector \(w\), with the addition of a bias term \(b\).
Each weight \(w_i\) directly relates to the input feature \(x_i\) and characterizes how relevant the \(i\)-th dimension is for discriminating the classes.
As a further contribution, \(b\) spatially shifts the position of the decision boundary away from the origin. 
Lastly, LR employs the sigmoid function \(S(x) = 1 / (1 - exp(-x))\) to map real-valued numbers into the range \([0,1]\), thus retrieving the class probability. 

To support multi-class classification, we leverage the one-vs-all approach, which consists of training \(N_{class}\) distinct binary classifiers, each designed to recognize a specific class against the others.
Thus, the learned vector \(W\) becomes a matrix of size \(N_{class}\,\times\,d\), while \(b\) is a \(N_{class}\) dimensional vector.
Each classifier output is a real value representing the predicted score of the target class.
The Softmax function shown in Eq.~(\ref{eq:softmax}) normalizes the result to a probability distribution over the output classes.
Lastly, the ArgMax operator (\ref{eq:argmax}) selects the class characterized by the largest predicted probability.
\begin{equation}
    f(x) = S(wx + b)\label{eq:lr_decision_function}
\end{equation}
\begin{equation}
    \sigma(x_{i}) = \frac{\exp(x_i)}{\sum_j \exp(x_j)},\,\,\,i\in[0,N_{class}-1]   \label{eq:softmax}
\end{equation}
\begin{equation}
    y = {\mathrm{ArgMax}}\,[\sigma(Wx + b)] \label{eq:argmax} 
\end{equation}
\subsubsection{Support Vector Machine (SVM)}
\label{SVM}
SVM is a linear ML model that provides a robust theoretical foundation and generalization performance \cite{cortes1995svm}.
Several domain-specific applications rely on SVM due to its ability to handle high-dimensional data and solve non-linear tasks.
Yi-Hung et al. \cite{svm2007yi} proposed an SVM-based face recognition system, while Siddharth et al. \cite{svm2020sid} introduced an EEG-based focal seizure detection algorithm that deploys SVM with 100\% accuracy.

In the binary classification setting, SVM consists of an optimal \((d-1)\) dimensional hyperplane determined by the \(d\)-dimensional normal vector \(w\) and the offset \(b\) that separates the training set \(A\) into classes by the largest margin.
The nearest data points to the hyperplane represent the Support Vectors (SVs), while their distance corresponds to the margin.
Although the general formulation of the algorithm enables classifying non-linearly separable data via high-dimensional mapping, we only focus on a linear kernel in this work.

SVM inference involves processing \(x\) deploying the decision function described in Eq.~(\ref{eq:svm_decision_function}), where \(sign\) refers to the function extracting the argument sign.
Thus, \(wx + b\) indicates on which side of the generated hyperplane the testing input \(x\) resides, while the \(sign\) function extrapolates the information providing the output class. 
Moving towards multi-class configuration, we leverage the one-vs-all approach again, learning a hyperplane per class.
\begin{equation}
    y = sign(wx + b)\label{eq:svm_decision_function}
\end{equation}

\begin{figure}[t]
    \centering
    \captionsetup{justification=centering}
    \includegraphics[width=0.6\linewidth]{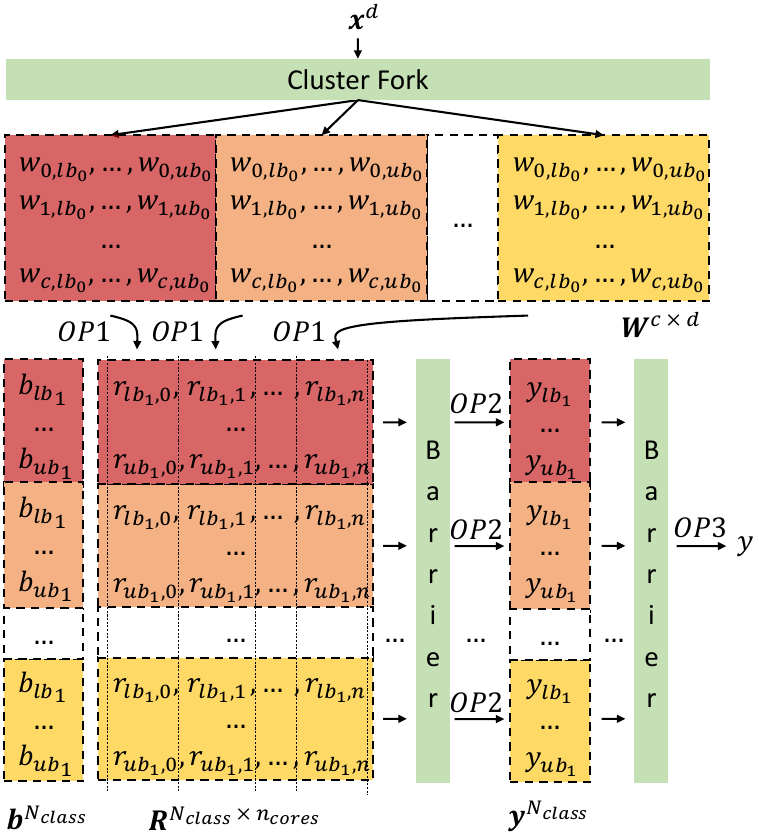}
    \caption{\REVA{GEMM-based Algorithms Parallelization Scheme}\\
    \textmd{
    \(OP1\): Partial matrix-vector multiplication,
    \(OP2\): Intermediate results and bias combination, \\
    \(OP3\): Activation function + ArgMax, \(\textit{\textbf{b}}\): Bias vector, 
    \(\textit{\textbf{R}}\): Matrix-vector multiplication intermediate result matrix \\
    \(d\): Dimension, \(c=N_{class}-1\), \(n=n_{cores}-1\), \\
    \(chunk_0=d/n_{cores},\,lb_0=core_{id}\,\times\,chunk_0,\,ub_0=lb_0+chunk_0 \),\\
    \(chunk_1=N_{class}/n_{cores},\,lb_1=core_{id}\,\times\,chunk_1,\,ub_1=lb_1+chunk_1 \),\\
    }
    }
    \label{fig:GEMM_PAR}
\end{figure}
\subsubsection{GEMM-based algorithms parallelization scheme}
In Figure~\ref{fig:GEMM_PAR}, we present the parallel design of GEMM-based algorithms optimized to maximize the speedup running on multi-core shared-memory platforms.
To offload the compute-intensive matrix-vector multiplication between \(\textit{\textbf{x}}\) and \(\textit{\textbf{W}}\) onto the CL, we assign to the cores the processing of \(chunk_0\) elements for each \(\textit{\textbf{W}}\) row following the \REVA{horizontal} decomposition scheme.
By using the offline determined \(chunk_0\) size and the \(core_{id}\), the cores compute at runtime lower (\(lb_0\)) and upper bounds (\(ub_0\)) data indexes for the first computation.
\(OP1\) consists of a partial matrix-vector multiplication where each core processes a \(\textit{\textbf{W}}\) row chunk multiplying and accumulating with the chunked input \(\textit{\textbf{x}}\).
Iterating the processing on \(\textit{\textbf{W}}\) rows, we store core-dependant intermediate results in a \(N_{class}\,\times\,n_{cores}\) sized shared global array \(\textit{\textbf{R}}\).
\REVA{After getting through a synchronization barrier, we obtain the effective matrix-vector multiplication result by combining intermediate results \(\textit{\textbf{R}}\) with vector \(\textit{\textbf{b}}\) and switching to a \REVA{vertical} parallel scheme in \(OP2\).}
Namely, the computation consists of accumulating \(\textit{\textbf{R}}\) elements by row with the corresponding \(\textit{\textbf{b}}\) value.
By leveraging a fresh \(chunk_1\), we calculate core-dependent \(lb_1\) and \(ub_1\) bounds which defines \(\textit{\textbf{b}}\) elements and \(\textit{\textbf{R}}\) rows assigned to each core. 
Thus, each core iterates on the \(chunk_1\) size accumulating \(\textit{\textbf{R}}\) rows with \(\textit{\textbf{b}}\) elements and leading to the \(N_{class}\) sized result vector \(\textit{\textbf{y}}\).
A CL synchronization barrier forces cores to wait until all CL cores finish \(OP2\) computation to avoid L1 data coherency issues. 
Consequently, the core master executes a sequential activation function \(OP3\) depending on the specific GEMM-based algorithm. 
LR requires the Softmax function to normalize the result, while SVM includes the \(sign\) routines to retrieve the argument sign. 
Lastly, \(OP3\) ends with the ArgMax to return the class with the highest score.

\subsection{Gaussian Naive Bayes (GNB)}
\label{GNB}
Naive Bayes (NB) consists of a family of simple probabilistic classifiers based on Bayes’ theorem along with the strong assumption of conditional independence among features given the class \cite{friedman1997nb}.
The model simplicity and high accuracy levels make the method attractive in several tasks, such as anomaly detection in industrial IoT \cite{gnb2020wu} and vehicle accident detection \cite{veh2021ku}.

Considering a multi-class problem while attempting to classify an input \(x\), the minimum classification error is ensured by picking the class \(c_i\) with the largest posterior probability \(P(c_i|x)\).
As shown in Eq. (\ref{eq:10}), Bayes' theorem enables to calculate posterior probabilities \(P(c_i|x)\) by leveraging prior probabilities \(P(c_i)\) and class-conditional likelihood \(P(x|c_i)\).
Since the marginal probability \(P(x)\) does not depend on the class \(c_i\) and \(x\) is constant, NB ignores \(P(x)\) calculation only keeping the joint probability \(P(x,c_i)\) in the numerator.
By using the chain rule to expand the definition of \(P(x,c_i)\) along with the strong conditional independence assumption, the joint probability model can be expressed as reported in Eq. (\ref{eq:11}).
\begin{equation}
    P(c_i|x) = \frac{P(x|c_i)P(c_i)}{P(x)} \propto P(x|c_i)P(c_i) = P(x,c_i),\,\,\,i\in[0,N_{class}-1] \label{eq:10}
\end{equation}
\begin{equation}
    P(c_i|x) \propto P(c_i) \prod_{k = 1}^{d-1} P(x_k|c_i),\,\,\,i\in[0,N_{class}-1] \label{eq:11}
\end{equation}

We derive the NB classifier by combining the model mentioned above and the Argmax decision rule (\ref{eq:12}).
\begin{equation}
    y = \underset{i\,\in\,N_{class}}{\mathrm{ArgMax}}\,P(c_i) \prod_{k = 1}^{d-1} P(x_k|c_i) \label{eq:12}
\end{equation}
NB classifiers differ mainly by the assumptions made regarding the distribution of the class-conditional likelihood \(P(x|c_i)\).
In this work, we leverage a normal Gaussian distribution (\ref{eq:13}) to estimate statistical parameters for features.
By performing a Maximum-Likelihood training, we learn the \(N_{class}\,\times\,d\) sized mean (\(\mu\)) and variance (\(\sigma\)) matrices, while the \(N_{class}\) dimensional prior probability \(P(c_i)\) vector is estimated directly on the dataset.
\begin{equation}
    P(x|c_i) = \frac{1}{\sqrt{2\pi\sigma^{2}_{i}}} \exp\left( -\frac{(x-\mu_{i})^2}{2\sigma^{2}_{i}}\,\right),\,\,\,i\in[0,N_{class}-1]
    \label{eq:13}
\end{equation}

\begin{figure}[t]
    \centering
    \captionsetup{justification=centering}
    \includegraphics[width=1\linewidth]{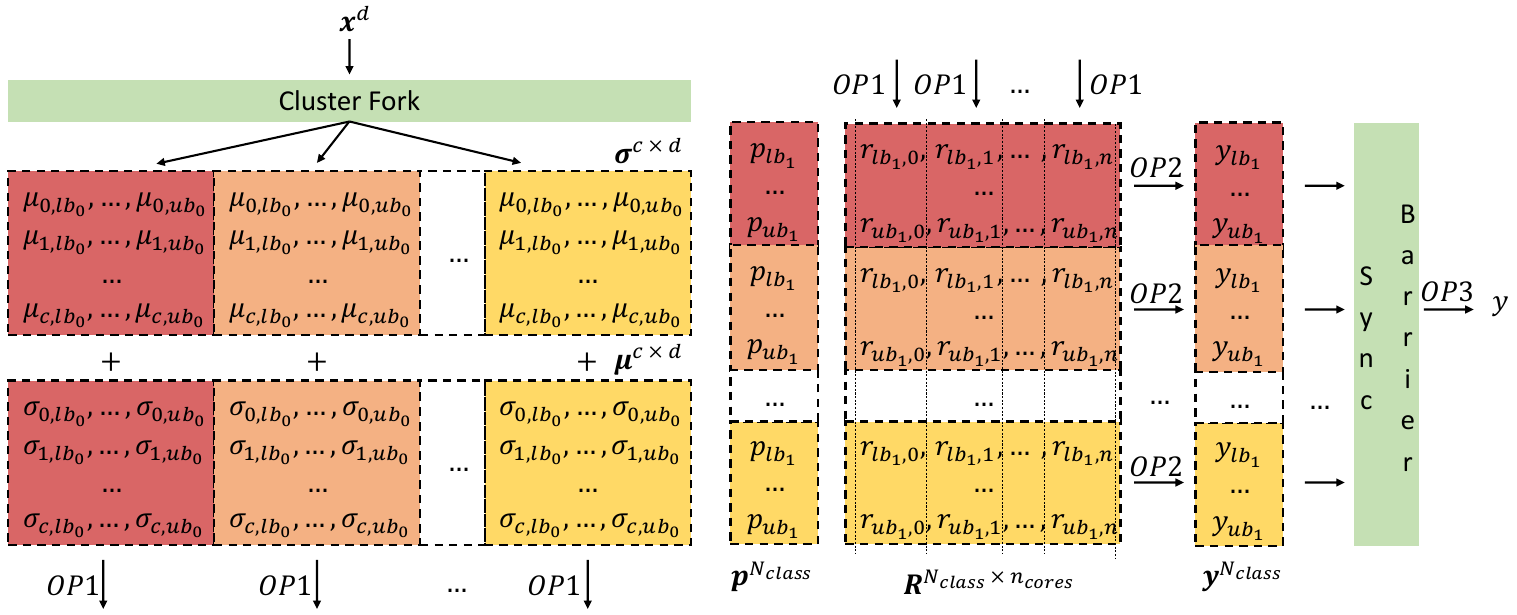}
    \caption{\REVA{GNB Parallelization Scheme}\\
    \textmd{
    \(OP1\): Partial \(P(x|c)\) sequence product, \(OP2\): Intermediate results and \(\textit{\textbf{p}}\) combination,\\
    \(OP3\): ArgMax, \(\textit{\textbf{p}}\): Prior probabilities vector, \(\textit{\textbf{R}}\): Sequence product intermediate result matrix \\
    \(d\): Dimension, \(c=N_{class}-1\), \(n=n_{cores}-1\) \\
    \(\,chunk_0=d/n_{cores},\,lb_0=core_{id}\,\times\,chunk_0,\,ub_0=lb_0+chunk_0 \) \\
    \(\,chunk_1=N_{class}/n_{cores},\,lb_1=core_{id}\,\times\,chunk_1,\,ub_1=lb_1+chunk_1 \)
    }
    }
    \label{fig:GNB_PAR_1}
\end{figure}
\subsubsection{GNB parallelization scheme}
To perform NB decision function (\ref{eq:12}) while fully leveraging CL compute power, we designed the parallelization scheme shown in Figure~\ref{fig:GNB_PAR_1}.
GNB per-class key operation consists of computing feature-dependant class-conditional likelihoods \(P(x_k|c_i)\) and combining them in a sequence product with the prior probability \(P(c_i)\). 
In \(OP1\), we vertically split this compute-intensive workload, assigning each CL core a partial sequence product by leveraging an optimal \(chunk_0\) data size computed offline.
At runtime, each core calculates core-dependent \(lb_0\) and \(ub_0\) data index boundaries to retrieve \(chunk_0\) per-row \(\boldsymbol{\mu}\) and \(\boldsymbol{\sigma}\) elements necessary to compute \(P(x_k|c_i)\).
By applying the Gaussian distribution formula (\ref{eq:13}) for each \(\mu - \sigma\) pair in the core-dependent \(chunk_0\) and multiplying them, we place \(OP1\) results in an intermediate \(N_{class}\,\times\,n_{cores}\) sized shared array \(\textit{\textbf{R}}\).
To bring together intermediate results and achieve the actual result, we combine \(\textit{\textbf{R}}\) with \(\textit{\textbf{p}}\) vector in \(OP2\) by leveraging a \REVA{vertical} decomposition scheme.
Thus, we define at compile time a fresh \(chunk_1\) data size, determining the number of \(\textit{\textbf{p}}\) elements and \(\textit{\textbf{R}}\) rows assigned to each core. 
By calculating \(lb_1\) and \(ub_1\) bounds, the cores iterate vertically on \(chunk_1\) rows multiplying \(\textit{\textbf{p}}\) with core-related partial sequence product and resulting in the \(N_{class}\) sized result vector \(\textit{\textbf{y}}\). 
Since \(OP3\) consists of a sequential computation on \(\textit{\textbf{y}}\), we deploy a CL synchronization barrier to force waiting until all CL cores finish \(OP2\) operation.
Lastly, the core master retrieves the class \(y\) with the highest score by performing the ArgMax function.

\subsection{Metric Space based Algorithms}
\label{Metric_Space_based_Algorithms}
MS-based algorithms involve arranging data points by proximity order leveraging the computed distances.
In this work, we consider the Euclidean metric shown in Eq. \ref{eq:5}.
In addition, we provide a time complexity analysis on alternative sorting algorithms when running on a sequential and parallel platform, respectively.
\begin{equation}
    \lVert p - q \rVert = \sqrt{\sum_{i=1}^{d-1} (p_{i} - q_{i})^2}\label{eq:5}   
\end{equation}
\subsubsection{k-Nearest Neighbor (kNN)}
\label{kNN}
kNN is a non-parametric instance-based supervised learning algorithm widely used in classification problems \cite{cover1967knn}. 
Due to its simplicity and classification performance, the model has been adopted in gesture recognition ML systems \cite{gest2021wo} and bone cancer detection approaches \cite{bone2019ran}. 

Without learning a discriminative function from the training set  \(A\), kNN stores the whole set and delays computations until inference.
Given a testing input \(x\) and a distance function, kNN computes the distance between \(x\) and \(A\).
The model orders \(A\) instances in descending order of proximity through the retrieved distances. Finally, kNN classifies \(x\) as the most prevalent class among the \(k\) nearest neighbors to the query point.

\subsubsection{\(k\)-Means}
\label{k_means}
\(k\)-Means \cite{macqueen1997kmeans} is a well-known unsupervised learning algorithm widely deployed in several domains, such as data mining \cite{data2017wu} and pattern recognition \cite{pattern2013peng}. Without requiring a training phase, the clustering method relies on an iterative pass that partitions the training set \(A\) space into disjointed regions covering the original input space.
Considering dividing \(A\) into \(k\) clusters \(U_{j\in[0,\,k-1]}\), each represented by arbitrarily initialized \(d\)-dimensional centroids \(u_{j\in[0,\,k-1]}\), the iterative procedure consists of the following steps:

\begin{itemize}
    \item Distance calculation: compute the Euclidean distance \(\lVert p - q \rVert\) between \(A\) and clusters centroids \(u_j\), as indicated in Eq. (\ref{eq:9}).
\begin{equation}
    d_{j\,+\,k\,\times\,i} = \lVert x_i - u_j \rVert \qquad j\in[0,k-1],\ i\in[0,N_{train}-1] \label{eq:9}   
\end{equation}
    \item Clusters allocation: assign data instances to the nearest centroid \(u_j\) according to Eq. (\ref{eq:6}), where \(i\) represents the \(i\)-th \(A\) instance and \(id_i\) the assigned cluster.
\begin{equation}
    id_i = \arg\min_{} d_{j\,+\,k\,\times\,i} \qquad j\in[0,k-1],\ i\in[0,N_{train}-1] \label{eq:6}   
\end{equation}
    \item Centroids update: compute new centroid \(u^{new}_j\) coordinates by averaging the instances belonging to the corresponding cluster \(u^{old}_j\), as reported in Eq. (\ref{eq:8}).  
\begin{equation}
    u^{new}_j = \frac{\sum_{i = 0}^{N - 1} I\{id_i = j\}\ x_i}{\sum_{i = 0}^{N - 1} I\{id_i = j\}} \qquad j\in[0,k-1] \label{eq:8}   
\end{equation}
\end{itemize}
\(k\)-Means continues iterating the three steps until the distance between previous \(u^{old}_j\) and current centroids \(u^{new}_j\) is lower than a pre-fixed threshold. 
When the centroids do not move significantly between iterations, the algorithm reaches the final centroids. 
In this work, we pick the first \(k\) elements of the training set \(A\) as initial centroids for \(k\)-Means clusters.
\subsubsection{Sorting Algorithms}
\label{sorting_algorithms}
MS-based algorithms require arranging data points based on the computed distances.
Traditional efficient sorting routines feature a favorable time complexity when dealing with complete sorting problems.
By the way, kNN and k-Means demand a partial sort returning the \(k\) smallest elements and the smallest one, respectively.
Considering a \(n\)-sized input array, retrieving the lowest \(k\) elements without sorting the remaining \(n - k\) elements could lead to a significant speedup improvement.
For that purpose, we present a brief time-complexity analysis of two well-known sorting routines, highlighting the advantages and drawbacks when running on a sequential and parallel platform.

Quick Sort (QS) is a highly efficient in-place sorting algorithm based on a divide-and-conquer procedure.
By selecting a pivot element, the routine partitions the input array into two sub-arrays and reorders them, relying on the pivot comparison.
The procedure is then re-iterated recursively on the sub-arrays until obtaining the reordered input array.
QS routine has a time complexity of \(O(n\log_2{n})\) on average when executing on a single-core platform.
Due to the divide-and-conquer algorithm nature, QS complexity does not scale when dealing with a partial sorting task.
Thus, the routine requires ordering the whole input array making its adoption highly inefficient for MS-based algorithms.

Selection Sort (SS) is a simple in-place comparison-based sorting algorithm that separates the input array into two sub-arrays. 
Initially, the sorted sub-array is empty, while the unsorted sub-array consists of the whole input array. 
By finding the smaller element in the unsorted sub-array, the algorithm swaps it with the leftmost unsorted element and moves the sub-array boundaries. 
Although the SS procedure offers the worst time complexity on average (\(O(n^2)\)), it enables saving computations when tackling partial sorting problems.
Considering returning the \(k\) smallest element, SS demands \(O(nk)\) comparisons, making its adoption in MS-based algorithms favorable compared to QS when \(k<\log_2{n}\). 
Deploying SS with k-Means is highly efficient since the algorithm determines the closest centroid for each data instance, corresponding to \(k = 1\).
Regarding kNN, the most efficient sorting algorithm strictly depends on the dataset dimension \(n\) and the hyperparameter \(k\). 
In this work, we deploy for kNN and k-Means a dataset consisting of 1k instances, favoring SS deployment when \(k < 10\).

When moving to a multi-core CL composed of \(c\) cores, the operating array is divided into \(c\) sub-arrays.
Each core performs the sorting routine on the corresponding local sub-array requiring \(O(\frac{n}{c}\log_2{(\frac{n}{c})})\) and \(O(\frac{n}{c}k)\) comparisons for QS and SS, respectively.
To bring together local results, an additional set of comparisons between the local smaller \(k\) elements is mandatory, requiring \(O(ck)\) comparisons.
In Eq. \ref{eq:9}, we report the time complexity of the two sorting algorithms, noting that the parallelization introduces an equal overhead on both routines.
Thus, running on a multi-core platform makes SS adoption favorable compared to QS when \(k<\log_2{(\frac{n}{c})}\).
As in the sequential execution, SS is still highly efficient in k-Means, while in kNN, the hyperparameter \(k\) determines the most efficient sorting algorithm.
Considering the 1k instances dataset used for kNN and k-Means, SS is favorable when \(k < 7\).
\begin{equation}
    QS = O(\frac{n}{c}\log_2{(\frac{n}{c})}) + O(ck)\;\;\;\:
    SS = O(\frac{n}{c}k) + O(ck)
    \label{eq:}
\end{equation}

\begin{figure}[t]
    \centering
    \captionsetup{justification=centering}
    \includegraphics[width=0.7\linewidth]{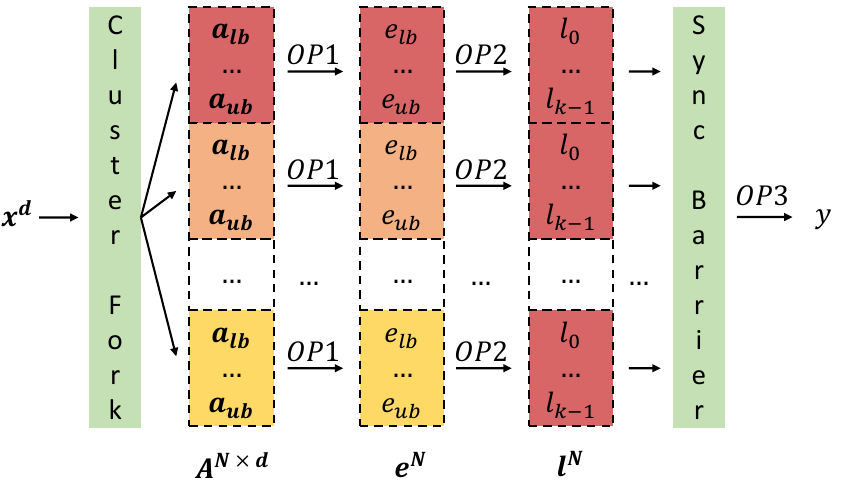}
    \caption{kNN Parallelization Approach\\
    \textmd{
    \(OP1\): Euclidean Distance, \(OP2\): k-elements Local Selection Sort,\\
    \(OP3\): k-elements Selection Global Sort + ArgMax, \(\textit{\textbf{A}}\): Training set, 
    \(\textit{\textbf{e}}\): Euclidean distance vector,\\
    \(\textit{\textbf{l}}\): Local k-nearest neighbors vector, \(d\): Dimension, \(k\): Nearest neighbors hyperparameter\\
    \(N=N_{train},\,chunk=N/n_{cores},\,lb=core_{id}\,\times\,chunk,\,ub=lb+chunk \)
    }
    }
    \label{fig:KNN_PAR}
\end{figure}
\subsubsection{MS-based algorithms parallelization}
\label{k_}
Figure~\ref{fig:KNN_PAR} shows the parallelization approach designed to dispatch kNN inference onto the 8-core CL.
The first operation (\(OP1\)) consists of computing the Euclidean distance between the query point \(\textit{\textbf{x}}\) and \(\textit{\textbf{A}}\), thus \(N_{train}\) distance operations.
To fully leverage the CL compute power, we use a \REVA{vertical} decomposition scheme to split the workload and determine offline the \(chunk\) size on which each core works.
At run-time, the cores calculate individual lower (\(lb\)) and upper bounds (\(ub\)) based on the \(core_{id}\) and perform the Euclidean distance computation on the corresponding \(chunk\) of \(\textit{\textbf{A}}\) rows.
After filling with results an intermediate \(N_{train}\) sized global array \(\textbf{e}\), the cores execute a \(k\)-elements Local Selection Sort (\(OP2\)) on the related \(chunk\), saving the local \(k\) neighbors in a \(N_{train}\)-dimensional global buffer \(\textit{\textbf{l}}\).
A CL synchronization barrier forces cores to wait until all CL cores finish \(OP2\) computation. To bring together intermediate results, the master core performs a \(k\)-elements Global Selection Sort (\(OP3\)) and returns the most voted class among the \(k\) neighbors performing the ArgMax function.

While kNN inference consists of a single procedure step, k-Means iterates a set of routines until the distance between \(\textit{\textbf{U}}_{new}\) and \(\textit{\textbf{U}}_{old}\) is smaller than a threshold.
In this regard, we present the optimized design of a k-Means iteration to achieve peak performance when running on a multi-core platform.

\begin{figure}[t]
    \centering
    \captionsetup{justification=centering}
    \includegraphics[width=0.8\linewidth]{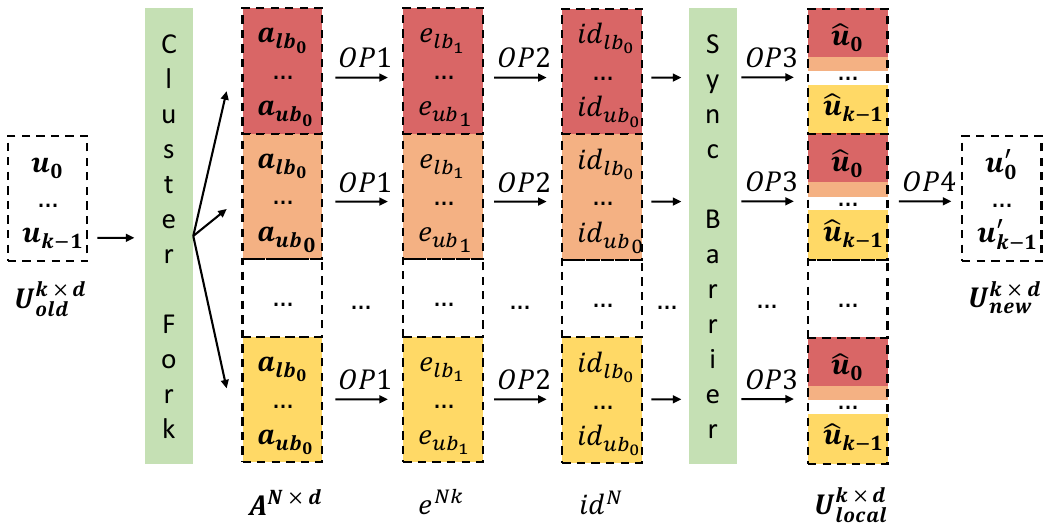}
    \caption{kmeans Parallelization Approach\\
    \textmd{
    \(OP1\): Euclidean distance calculation, \(OP2\): Cluster ID allocation, \(OP3\): Local \REVA{centroids} update,\\
    \(OP4\): Global \REVA{centroids} update, \(\textit{\textbf{A}}\): Training set, \(\textit{\textbf{e}}\): Euclidean distance vector, \(\textit{\textbf{id}}\): Cluster ID vector,\\
    \(\textit{\textbf{U}}_{old}\): Initial cluster centroids, \(\textit{\textbf{U}}_{local}\): Local cluster centroids, \(\textit{\textbf{U}}_{new}\): New cluster centroids,\\
    \(N = N_{train}\), \(chunk_0=N/n_{cores},\,lb_0=core_{id}\,\times\,chunk_0,\,ub_0=lb_0+chunk_0 \)\\
    \(chunk_1=(N\,\times\,k)/n_{cores},\,lb_1=core_{id}\,\times\,chunk_1,\,ub_1=lb_1+chunk_1 \)\\
    }
    }
    \label{fig:kmeans_PAR}
\end{figure}
As shown in Figure~\ref{fig:kmeans_PAR}, the algorithm begins calculating the Euclidean distance (\(OP1)\) between \(\textit{\textbf{A}}\) elements and each centroid \(\textit{\textbf{u}}_i\), thus demanding \(N \times k\) distance computations.
To dispatch the workload efficiently onto the CL, we divide \(\textit{\textbf{A}}\) horizontally by determining offline \(chunk_0\) which defines the number of \(\textit{\textbf{A}}\) rows assigned to each core.
At run-time, we offload the distance computation to each core using \(lb_0\) and \(ub_0\) to tag core-dependent data indexes.
Since a core computes \(k\) distances for each \(chunk_0\) element, \(OP1\) leads to a \(N \times k\) dimensional result that we store in the global shared buffer \(\textit{\textbf{e}}\).

In \(OP2\) the increased vertical dimension \((N \times k)\) demands expanding the data chunk to \(chunk_1\), making a core working on \(k\) distances for each \(chunk_0\) element.
Thus, the cores find the closest centroid \(\textit{\textbf{u}}_i\) to each \(chunk_0\) element and assign the cluster ID.
Furthermore, the results are saved in an \(N_{train}\)-sized array \(\textit{\textbf{id}}\) containing the cluster ID for each \(\textit{\textbf{A}}\) data sample.
\(OP3\) consists of a Local Centroids Update where each core accumulates and counts \(\textit{\textbf{A}}\) instances belonging to the same centroid \(\textit{\textbf{u}}_i\) operating on \(chunk_0\) elements. 
The operation ends with a CL synchronization barrier to ensure each core finishes the workload before moving to the following computation step.
Lastly, we perform a Global Centrodis Update (\(OP4\)) to pull together local results \(\textit{\textbf{U}}_{local}\).
Each core takes charge of computing the global value of a centroid \(\textit{\textbf{u}}_i\) corresponding to its \(core_{id}\), working on non-contiguous elements.
Thus, the core accumulates \(\textit{\textbf{U}}_{local}\) and count variables using the \(core_{id}\) to retrieve data from the chunks and dividing them, finds the new global centroid \(\textit{\textbf{U}}_{new}\).

\subsection{Random Forest}
\label{RF}
RF is a robust ML algorithm leveraging an ensemble of low-correlated randomized Decision Trees (DTs) to split the training set using feature space subsets \cite{breiman2001rf}. 
Due to the low-variance nature and the capability to handle various data types effectively, the model has been largely deployed in several domain-specific applications such as Non-Intrusive Load Monitoring \cite{tabanelli2020feature} and anomaly detection \cite{ad2020lin}. 

Starting from the root node, DTs consist of several splitting nodes where an input feature \(x_i\) is evaluated with a test condition to determine the branch to be followed.
Repeating the decision procedure over the entire structure, the DT reaches a leaf containing the predicted class.
Lastly, RF returns the input prediction by aggregating DTs votes and picking up the class with the higher number of votes.

To optimize the model execution on edge devices, we designed a custom DT implementation representing the model structure with arrays.
This approach save all tree structures into four arrays: feature, threshold, left child, and right child.
By using feature and threshold arrays, we evaluate the node comparison.
While leveraging the result, we pick the following node from the left- and right-child array.
Lastly, we mark leaf nodes by writing a negative integer value in the corresponding \(i\)-th node elements of the feature array.
\subsubsection{RF Parallelization Approach}
The DT algorithmic structure prevents a priori knowledge of the taken pathway toward the leaf at compile time.
The model unveils the taken branches by evaluating the input \(x\) at runtime, and this unpredictability complicates the DT parallelization.
In this regard, we adopt a parallelization scheme consisting of assigning the whole DT execution to a specific core. 
Furthermore, the strategy involves the static assignment of DTs to the available cores.

\begin{figure}[t]
    \centering
    \captionsetup{justification=centering}
    \includegraphics[width=0.85\linewidth]{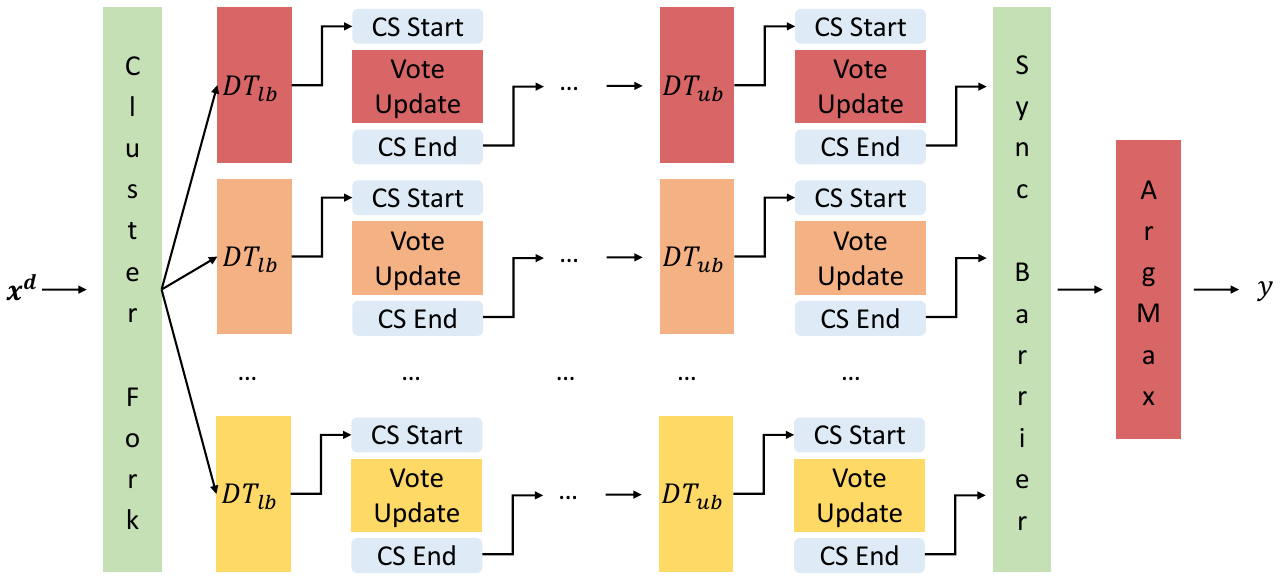}
    \caption{RF Parallelization Approach\\
    \textmd{
    \(DT_i\): i-th Decision Tree,
    CS: Critical Section, \(d\): Dimension\\
    \(chunk=N_{trees}/n_{cores}, \, lb=core_{id}\,\times\,chunk, \, ub=lb+chunk\)
    }}
    \label{fig:rf_PAR}
\end{figure}
In Figure~\ref{fig:rf_PAR}, we illustrate the parallel algorithm design to offload RF execution onto multi-core platforms maximizing the compute power utilization.
To efficiently dispatch the RF model onto the CL, we determine offline a \(chunk\) size representing the number of DTs assigned to each core. 
By computing core-dependant \(lb\) and \(ub\), each core retrieves the assigned \(DT_{id}\) and executes the workload computing the result for the assigned DTs.
A Critical Section (CS) barrier prevents multiple cores from accessing the Vote Update section simultaneously.
Thus, we aggregate DTs results atomically by incrementing the retrieved class in a vote array.
Lastly, a CL Synchronization Barrier ensures that each core finishes the workload before moving to the ArgMax function, which retrieves the final prediction.

%% file: 05-experiments.tex
\section{Experimental Evaluation}
This section presents the results of our design optimized for parallel execution employing a fine-grained analysis and intensive optimization. 
We provide Non-Neural ML algorithms execution time, considering two alternative FP emulation libraries and FPU-native support. 
By comparing the kernel single-core execution, we point out the performance improvement obtained by switching from a standard to a custom RISCV-based emulation support and an FPU-native platform.
We also compare achieved speedups for each target platform leveraging the 8-core CL compute power and the optimized algorithm parallel design.
To clarify the achieved results, we conducted an analysis to determine non-ideality sources and architectural factors when performance is sub-optimal.

Section~\ref{Setup} describes the adopted experimental setup and the ML framework deployed to train the Non-Neural ML kernels. 
A comparison of the sequential execution overhead between alternative FP emulation supports and an FPU-native platform is discussed in Section~\ref{Seq_FP_Support}.
After presenting in Section~\ref{Parallel_Performance} the achieved speedups by fully exploiting the CL compute power, we illustrate an in-depth comparison of the execution time between PULP-OPEN and ARM Cortex-M4 in Section~\ref{ARM}.

\subsection{Setup}
\label{Setup}
The experimental analysis has been conducted using two different target platforms. 
The GAPUINO development board\footnote{\url{https://greenwaves-technologies.com/product/gapuino/}} represents a commercial solution integrating GAP8 coupled with a rich set of peripheral interfaces to fast prototype embedded applications. 
A JTAG bridge allows programming the onboard FLASH memory and debugging GAP8 code.
Instead, the hardware design includes a set of Special-Purpose Registers (SPRs) to store the count of hardware-related events at the core level. 
Using non-intrusive per-core performance counters enables fine-grained performance analyses, measuring events related to instructions (executed instructions, total and active cycles) and memory accesses (I\$ misses, TCDM contentions, and L2/TCDM memory stalls).
In this work, we use the GAPUINO board to profile Non-Neural ML algorithms performance on GAP8 while using a standard and a custom software FP library. 
Furthermore, we set the FC clock frequency to 250MHz while the CL runs at 150MHz. 

We also performed experiments on the PULP-OPEN architecture, thus leveraging FPU-native support. 
To emulate the microarchitecture, we used a hardware emulator running on a Xilinx UltraScale+ VCU118 FPGA board\footnote{\url{https://www.xilinx.com/products/boards-and-kits/vcu118.html}}.
The architecture emulation enables faster experiments than RTL-equivalent simulations while providing cycle-accurate results. 
In addition to the performance counters provided by GAP8, the PULP-OPEN design supports recording FPU pipeline-related events (FPU stalls, contentions, and write-back stalls).
Using Vivado Design Suite, we generate and load the microarchitecture bitstream on the FPGA. An OpenOCD interface with GDB support mapped on GPIO pins allows uploading the application binary code in the L2 memory and running the program.
A virtual UART mapped on a dedicated USB port enables to read results from an emulated terminal. 
In this work, the FPGA clock frequency has been set to 20 MHz. 


\REVA{
To characterize performance, we selected three datasets widely adopted among the TinyML community and are contained in the MLPerf Tiny benchmark suite~\cite{banbury2021mlperf}.
Speech Commands is an audio dataset of spoken words designed to build Keyword Spotting systems, consisting of 105k utterances from 2.6k different speakers.
The dataset supports 35 English words and a collection of background noises, where each speech sample is 1sec long. 
Following MLPerf Tiny reference implementation, we deployed a subset of the dataset consisting of 10 words.
We used the remaining words to approximate the "unknown" label, which, along with "silence", results in 12 output classes. 
As pre-processing, we used 10 Mel-frequency cepstral coefficients (MFCC) features extracted from a 40~msec long speech frame with a stride of 20~ms, resulting in 490 features for 1sec audio. 
For that purpose, we used Speech Commands to benchmark GEMM-based algorithms in this work. 
To test MS-based algorithms, we deployed the ToyADMOS dataset for anomaly detection in machine operating sounds. 
According to MLPerf Tiny benchmark suite, we used only the Toy-car machine type among the other six available. 
For training, we deployed 7k normal sound samples from seven Toy-cars, each delivering 1k machine sound samples mixed with environmental noise. 
We also pre-processed the audio into a log-mel-spectrogram with 128 bands featuring a sliding window of five frames, leading to a 640 input size. 
Regarding k-Means, we adopted two 640-dimensional clusters to divide the training set, while four nearest neighbors for kNN. 
CIFAR-10 is a multi-class labeled dataset consisting of 60k 32x32 RGB images, divided into 50k training instances and 10k for the testing set. 
The dataset represents the de-facto standard for TinyML benchmarking since the low image resolution makes CIFAR-10 the most suited data source for training tiny image classification models.
For that purpose, we used CIFAR-10 to benchmark the IT-based algorithm and GNB in this work.
}

We performed the training of the algorithms entirely relying on the Scikit-Learn ML framework, leveraging its front-end to dump model parameters and structures. 
Whenever model parameters do not fit the L1 memory, we place data into the L2 level and use the double-buffering wrapper to overlap DMA operations with kernel processing optimally.
\REVA{To guarantee efficient runtimes, we initially optimized the sequential version of the Non-Neural ML algorithms on each platform. 
Thus, we thoroughly investigated kernel execution using non-intrusive performance counters to optimize the instruction-level scheduling of the 4-stage in-order single-issue pipeline adopted by both target cores.
We used the L1 load stall counter to limit hazards due to data dependencies while monitoring branch stalls to minimize pipeline flushing.
We also leveraged the I\$ misses counter to investigate cache locality issues. 
This in-depth analysis led to the highest attainable CPU utilization achieving near-optimal Clock per Instruction (CPI) for most algorithms.
In the parallel version, we focused on reducing TCDM contentions to limit the wasting of cycles when multiple cores attempt to read data from the same memory block.
Furthermore, we optimized the use of parallel programming primitives to the bare minimum reduce synchronization overheads.
}
Lastly, we conducted extensive benchmarking considering all FP emulation supports and platforms, measuring the execution cycles and other statistics for each variant.


\begin{figure}[htp!]
    \centering
    \captionsetup{justification=centering}
    \caption{\REVA{Non-Neural ML algorithms cycles, latency, and energy on a single-core GAP8 and PULP-OPEN configuration}}
    \includegraphics[width=1\linewidth]{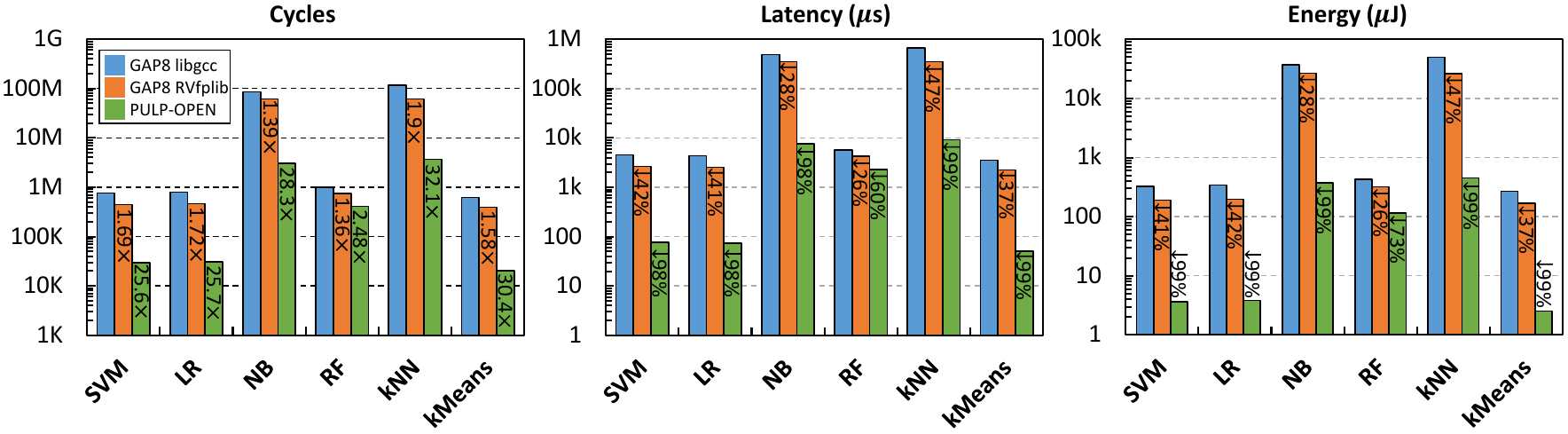}
    \label{fig:FP_EMUL}
\end{figure}
\subsection{Benchmarking Floating-Point Emulation Libraries vs FPU-Native Support}
\label{Seq_FP_Support}
%
%
\REVA{
In Figure~\ref{fig:FP_EMUL}, we show the cycles, latency, and energy required by Non-Neural ML algorithms considering a sequential execution on the two RISCV-based PULP MCUs and alternative FP emulation libraries for GAP8.
We report on top of cycles columns the achieved speedup compared to the baseline, which consists of executing the kernels on GAP8 with libgcc support for FP emulation. 
Regarding the energy efficiency and latency values, we indicate the percentage decrease compared to the baseline. 
Table~\ref{tab:fp_emul_code_size} represents algorithms code size and percentage reduction when moving from the baseline to RVfplib emulation and then to the FPU-native system. 
Lastly, we present in Table~\ref{tab:fp_emul_vs_fpu} the execution statistics for each kernel and platform configuration, along with the architectural non-idealities retrieved from the performance counters.
}
Pipeline Non-Idealities (N.I.) refers to the sum of architectural factors owed to the cores pipeline (stalls related to memory load latency and taken branches). At the same time, FPU N.I. accounts for FPU-related events limiting the efficiency (write-backs, contentions, and dependencies).
\REVA{
libgcc emulation leads to the lowest CPI, ranging from 1.28 to 1.45, due to the high usage of branching conditions and global variables placed into L2 memory by the GCC toolchain. 
Moving from the baseline to the custom RISCV-based RVfplib emulation library reduces execution times, achieving 1.36-1.9\(\times\) speedups on GAP8 and a higher 1.18-1.33 CPI.
Employing fast SW FP emulated routines on FPU-less processors brings several further benefits for TinyML: latency features up to 47.34\% decrease, while energy efficiency reaches 26.27\%-47.34\% reductions.
Adopting the FPU-native PULP-OPEN platform decreases pipeline N.I. and FPU factors to 1\% execution time, reaching up to 1.12 CPI and 32.09\(\times\) performance improvement compared to the baseline. 
Consequently, FP support leads to higher latency and energy lowering, ranging from 59.74\% to 99.1\% compared to libgcc adoption on top of GAP8.}
%
%
\begin{table*}[t]
    \centering
    \caption{\REVA{Runtime statistics and architectural factors executing the Non-Neural ML algorithms on a single-core GAP8 and PULP-OPEN configuration, leveraging libgcc and RVfplib for FP emulation on GAP8.}}
    \setlength\extrarowheight{1.5pt}
    \resizebox{\textwidth}{!}{
    \begin{tabular}{ccccccccccc} 
        \hlineB{2}
        Kernel & Platform & FP Instr. (\%) & Cycles & Instr. & \REVA{CPI} & Speedup & Pipeline N.I. & I\$ Misses & Ext. LD & FPU N.I\\
        \hline
        \hline
        \multirow{3}*{SVM}
            & GAP8 + libgcc & 89.98 & 757k & 548k & \REVA{1.38} & - & 146k & 7.6k & 4.5k & -\\
        \cline{2-11}
            & GAP8 + RVfplib & 69.06 & 447k & 335k & \REVA{1.33} & \(\mathbf{1.69}\) & 92.7k & 16.3k & 1 & - \\
        \cline{2-11}
            & PULP-OPEN & 24.89 & 29.6k & 23.7k & \REVA{1.25} & \(\mathbf{25.56}\) & 5.9k & 25 & 1 & 0 \\
        \hline
        \multirow{3}*{LR} 
            & GAP8 + libgcc & 90.16 & 796k & 570k & \REVA{1.40} & - & 150k & 24.8k & 4.60k & - \\
        \cline{2-11}
            & GAP8 + RVfplib & 68.65 & 463k & 351k & \REVA{1.32} & \(\mathbf{1.72}\) & 96.8k & 37 & 1 & - \\
        \cline{2-11}
            & PULP-OPEN & 24.98 & 30.9k & 24.6k & \REVA{1.26} & \(\mathbf{25.75}\) & 6.10k & 5 & 1 & 184 \\
        \hline
        \multirow{3}*{GNB} 
            & GAP8 + libgcc & 92.42 & 86.4M & 67.4M & \REVA{1.28} & - & 15.9M & 3.38M & 16.1k & - \\
        \cline{2-11}
            & GAP8 + RVfplib & 57.67 & 62.0M & 50.1M & \REVA{1.24} & \(\mathbf{1.39}\) & 11M & 387k & 1 & - \\
        \cline{2-11}
            & PULP-OPEN & 27.25 & 3.05M & 2.72M & \REVA{1.12} & \(\mathbf{28.34}\) & 279k & 37.9k & 1 & 30.7k \\ 
        \hline
        \multirow{3}*{RF}
            & GAP8 + libgcc & 54.23 & 1.01M & 695k & \REVA{1.45} & - & 344k & 39.9k & 1 & - \\
        \cline{2-11}
            & GAP8 + RVfplib & 29.98 & 742k & 629k & \REVA{1.18} & \(\mathbf{1.36}\) & 78.8k & 18.5k & 1 & - \\
        \cline{2-11}
            & PULP-OPEN & 6.39 & 405k & 350k & \REVA{1.16} & \(\mathbf{2.48}\) & 70.5k & 19.9k & 1 & 0 \\
        \hline
        \multirow{3}*{kNN}
            & GAP8 + libgcc & 90.49 & 117M & 80.7M & \REVA{1.45} & - & 29.1M & 1.57M & 554k & - \\
        \cline{2-11}
            & GAP8 + RVfplib & 69.68 & 61.6M & 46.5M & \REVA{1.32} & \(\mathbf{1.9}\) & 13.3M & 635k & 15 & - \\
        \cline{2-11}
            & PULP-OPEN & 45.5 & 3.64M & 2.85M & \REVA{1.28} & \(\mathbf{32.09}\) & 735k & 36.6k & 15 & 0\\
        \hline
        \multirow{3}*{kMEANS} 
            & GAP8 + libgcc & 74.82 & 625k & 466k & \REVA{1.34} & - & 89.4k & 8.39M & 515 & - \\
        \cline{2-11}
            & GAP8 + RVfplib & 48.27 & 395k & 315M & \REVA{1.25} & \(\mathbf{1.58}\) & 45.4k & 525 & 1 & - \\
        \cline{2-11}
            & PULP-OPEN & 40.64 & 20.5k & 18.3k & \REVA{1.26} & \(\mathbf{30.44}\) & 2.8k & 41 & 1 & 44\\
        \hlineB{2}
    \end{tabular}
    }
    \label{tab:fp_emul_vs_fpu}
\end{table*}

GEMM-based algorithms demand executing a matrix-vector multiplication, which requires a sequence of FP \textit{mul} and \textit{add} operations at low level.
When executing the kernel on the baseline, libgcc \_\_\texttt{mulsf3} and \_\_\texttt{addsf3} emulation routines (multiplication and addition between single-precision FP variables, respectively) slow down the runtime requiring about 800~kcycles per inference.
Compiling GAP8 code integrating the RISCV-based emulation library decreases the execution time to almost 450~kcycles due to the RVfplib latency obtained by leveraging the PULP ISA extensions.
Thanks to the native support for single-cycle FP arithmetic instructions, PULP-OPEN decreases further the execution time, leading to a 25.56-25.75\(\times\) speedup compared to the baseline.

In the GNB model, the normal Gaussian distribution calculation requires executing high latency transcendental functions (i.e., \texttt{expf} and \texttt{logf}), thus making the algorithm compute-intensive. 
As a result, running the kernel on the baseline setup demands an order of magnitude higher execution time than previous algorithms, namely 86.4~Mcycles.
By deploying RVfplib on GAP8, the executin time decreases to 62~Mcycles with a 0.3\(\times\) speedup drop compared to the performance of GEMM-based kernels. 
Transcendental functions involve a high usage of the \_\_\texttt{divsf3} routine, which slows down the execution time when passing from libgcc to RVfplib emulation support.
As a consequence, \texttt{expf} and \texttt{logf} routines present a 1.2\(\times\) average speedup with respect to the baseline. 
Overall, transcendental functions severely limit RVfplib speedup since they account for 20\% of GNB execution time.
Furthermore, taken branches (TBs) account for 17.78\% GNB computational time and decrease by up to 5\% less than GEMM-based kernels, thus limiting the runtime improvement.
Moving the execution onto PULP-OPEN further reduces the running time to 3.05~kcycles, thus reaching a 28.34\(\times\) speedup compared to the baseline.
Load stalls reduction to almost 0\% of the execution time enables a 3\(\times\) relative speedup increase compared to GEMM-based kernels, where load stalls represent 19\% of the computation time.

Due to limited usage of FP computations, RF presents lower performance when switching the FP emulation support and moving to an FPU-native platform.
On the baseline, RF demands about 1.01~Mcycles deploying only the \_\_\texttt{lesf2} libgcc emulation routine to compare feature values with thresholds. 
By showing 54.23\% FP instructions, RVfplib allows improving only a limited fraction of the workload, thus leading to 742~kcycles with a 1.36\(\times\) speedup compared to the baseline.
Leveraging the PULP-OPEN FPU reduces the execution time to about 405~kcycles with a reduced speedup of 2.38\(\times\) owing to a 6.39\% kernel FLOP intensity.

By running kNN on GAP8 deploying libgcc FP emulation support, the kernel requires 117~Mcycles per inference.
Since the algorithm leverages GEMM-based FP emulation routines with the addition of \_\_\texttt{subsf3}, achieving a 1.9\(\times\) speedup with RVfplib is mainly due to architectural factors.
While TBs increase by 2.01\% of the execution time in GEMM-based kernels, kNN presents a TBs decrease of almost 3\% of the computing time moving from libgcc to RVfplib.
Previous algorithms feature 24.89-27.25\% FP instructions, while kNN reaches up to 45.5\% due to 21.2M FP instructions out of a total of 46.5M instructions.
As a result, the kernel gains performance from leveraging more of the FPU compute power leading to a 32.09\(\times\) speedup compared to the baseline when deploying PULP-OPEN.

The kernel takes about 625~kcycles when performing on the baseline while leveraging RVfplib on GAP8 reaches a 1.58\(\times\) speedup reducing the runtime to 395~Mcycles.
kMEANS lower FP rate compared to kNN explains the 0.3\(\times\) drop of performance when switching from libgcc to RVfplib FP support.
While kNN accounts for 90.49\% instructions to emulate FP computations, kMEANS uses only 74.82\% of the overall workload, thus leading to a speedup decrease.
Running the kernel on PULP-OPEN, the execution time decreases to almost 20.5~kcycles, improving performance by 30.44\(\times\) compared to the baseline.
By presenting a reduced FLOP intensity of 40.64\% and a higher LD stalls increase compared to kNN, the kernel achieves a 2\(\times\) lower speedup compared to the baseline. 

\REVA{
Adopting SW-optimized FP emulation libraries on IoT FPU-less platforms leads to several advantages also for latency and energy efficiency. 
GEMM- and MS-based algorithms are almost dominated by FP computations, featuring 75\% to 90\% of FP instructions. 
Leveraging small optimized RBfplib routines leads to 36.7\%-47.34\% energy usage reduction, demanding about 190~\(\mu\)J per GEMM-based and k-Means inference and 26.4~\(m\)J for kNN.
Consequently, such Non-Neural ML kernels present higher latency percentage reductions, enabling running inferences on GAP8 in about 352~\(m\)s for kNN and 2.5~\(p\)s for the remaining. 
GNB transcendental routines high usage and RF reduced FP computations ratio limit energy and latency improvements to 26.2\%-28.8\% compared to libgcc deployment on GAP8. 
Instead, leveraging PULP-OPEN FPU-native support reduces such resources by up to 99\%, requiring down to 3.7~\(\mu\)J and 75~\(\mu\)s per GEMM-based inference.
}

\REVA{
Adopting RVfplib on GAP8 to execute RF reduces the code size by only 3.9\% due to the low FP computations ratio, while the other kernels reach a 7.9\% lowering. 
Lastly, PULP-OPEN FPU-native support decreases the code size up to 42\%, considering libgcc support.
}


\begin{table*}[htp!]
    \centering
    \caption{\REVA{Non-Neural ML kernels code size on a single-core GAP8 and PULP-OPEN configuration, leveraging libgcc and RVfplib for FP emulation on GAP8.}}
    \resizebox{\textwidth}{!}{
    \begin{tabular}{ccccccccc} 
        \hlineB{2}
        & SVM & LR & GNB & RF & kNN & k-Means \\
        \hline
        \hline
        GAP8+libgcc & 21.4 & 23.11 & 25.59 & 21.22 & 23.17 & 22.9 \\
        GAP8+RVfplib & 19.9kB ($\downarrow$7.3\%) & 21.3kB ($\downarrow$7.9\%) & 23.7kB  ($\downarrow$7.3\%) & 20.4kB ($\downarrow$3.9\%) & 21.5kB ($\downarrow$7\%) & 21.3kB  ($\downarrow$7\%) \\
        PULP-OPEN & 13kB ($\downarrow$39\%) & 13.5kB ($\downarrow$42\%) & 15.4kB ($\downarrow$40\%) & 13kB ($\downarrow$39\%) & 14.2kB ($\downarrow$39\%) & 13.8kB ($\downarrow$40\%) \\
        \hlineB{2}
    \end{tabular}
    }
    \label{tab:fp_emul_code_size}
\end{table*}

\subsection{Parallel performance}
\label{Parallel_Performance}
\begin{figure}[htp!]
    \centering
    \captionsetup{justification=centering}
    \includegraphics[width=1\linewidth]{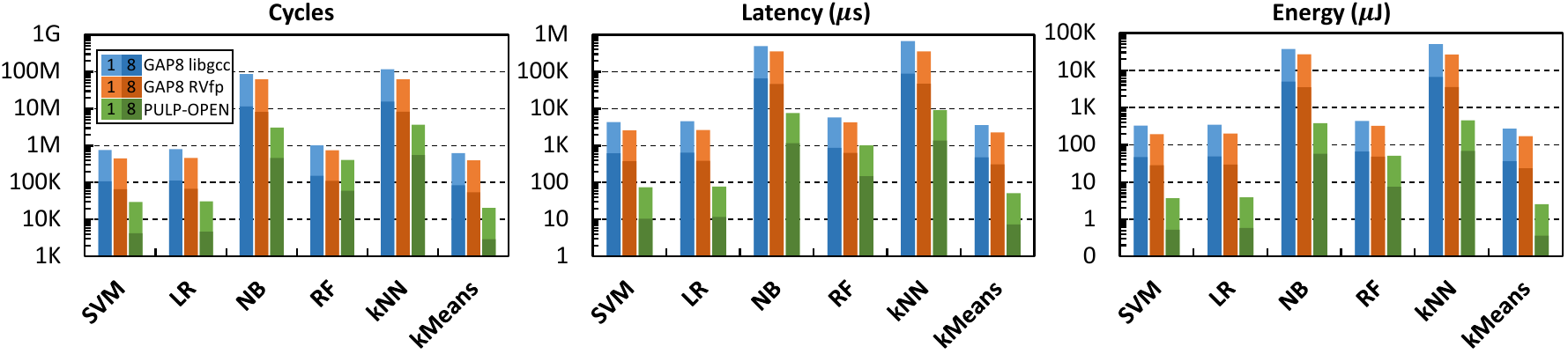}
    \caption{\REVA{1-vs-8 core Non-Neural ML algorithms cycles, latency, and energy comparison.\\
    Abbreviations: RVfp (RVfplib).}}
    \label{fig:ParallelPerformance_cycles_latency_energy}
\end{figure}

\begin{figure}[h!]
    \centering
    \captionsetup{justification=centering}
    \includegraphics[width=0.5\linewidth]{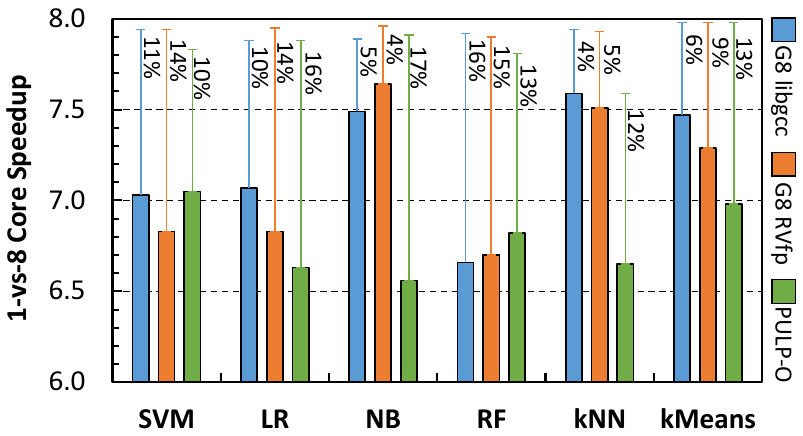}
    \caption{Non-Neural ML kernels parallel performance on GAP8 and PULP-OPEN.\\
    Abbreviations: G8 (GAP8), RVfp (RVfplib), PULP-O (PULP-OPEN).}
    \label{fig:PAR_SPEEDUP}
\end{figure}

%
%
\REVA{In Figure~\ref{fig:ParallelPerformance_cycles_latency_energy}, we report the cycles, latency, and energy required by Non-Neural ML kernels, comparing sequential and parallel execution on PULP-OPEN and GAP8.
To assess the parallelization performances, we also report the 1-vs-8 cores parallel speedup in Figure~\ref{fig:PAR_SPEEDUP} and indicate the percentage loss between the achieved and ideal speedup on top of each column.
Furthermore, Table~\ref{tab:parallel_performance} gives more profound insight into the results by providing measurements of the architectural factors limiting the speedup retrieved from platform performance counters.}
%
%
The considered ML kernels consist of a workload divided into fully parallelizable sections and inherently sequential portions.
For that purpose, the table also reports the theoretical speedup of Non-Neural ML kernels when using multiple processors.
Thus, we profiled the execution time of the sequential code sections for each platform configuration and applied Amdahl’s law using the formula in Eq.~(\ref{Amdahl}).
\begin{equation}
    Speedup = \frac{1}{(1-p) + \frac{p}{N}} \label{Amdahl}
\end{equation}
Amdahl’s law has two parameters: $p$ is the percentage of parallelizable code, and $N$ is the total number of available cores.
This formula provides an ideal bound for the theoretical speedup since it does not take into account the parallelization overheads. 
\REVA{
The optimized parallel design introduced in Section~\ref{Algorithm_Design} enables reaching near-ideal speedups ranging from 6.56\(\times\) to 7.64\(\times\) compared to a single-core execution.
By reducing TCDM contentions to at most 4.25\% of the execution time and improving the instruction scheduling, we achieve CPIs ranging from 1.32 to 1.72.}

\begin{table*}[t]
    \centering
    \caption{\REVA{Runtime statistics and architectural factors executing the Non-Neural ML algorithms on a single-core and 8-core configuration.}}
    \setlength\extrarowheight{1.5pt}
    \resizebox{\textwidth}{!}{
    \begin{tabular}{ccccccccccccc} 
        \hlineB{2}
        Kernel & Platform & Cores & Cycles & Instr. & \REVA{CPI} & Speedup & \makecell{Theor.\\Speedup} & \makecell{Pipeline\\N.I.} & \makecell{I\$\\Misses} & TCDM & \makecell{Ext.\\LD} & \makecell{FPU\\N.I.} \\
        \hline 
        \hline
        
        \multirow{6}*{SVM} 
            & \multirow{2}*{GAP8 + libgcc}
            & 1 & 757k & 548k & \REVA{1.38} & - & - & 146k & 7.6k & 0 & 4.5k & -\\
        \cline{3-13}
            & & 8 & 108k & 62.6k & \REVA{1.72} & 7.03 & 7.94 & 19.7k & 4.86k & 11 & 567 & - \\
        \cline{2-13}
            & \multirow{2}*{GAP8 + RVfplib}
            & 1 & 447k & 335k & \REVA{1.33} & - & - & 92.7k & 16.3k & 0 & 1 & - \\
        \cline{3-13}
            & & 8 & 65.5k & 45.2k & \REVA{1.45} & 6.83 & 7.94 & 12.3k & 2.67k & 16 & 2 & - \\
        \cline{2-13} 
            & \multirow{2}*{PULP-OPEN}
            & 1 & 29.6k & 23.7k & \REVA{1.25} & - & - & 5.9k & 25 & 0 & 1 & 0 \\
        \cline{3-13}
            & & 8 & 4.20k & 3.17k & \REVA{1.32} & 7.05 & 7.83 & 740 & 46 & 165 & 2 & 4 \\
        \hline
        \hline
        
        \multirow{6}*{LR} 
            & \multirow{2}*{GAP8 + libgcc}
            & 1 & 796k & 570k & \REVA{1.4} & - & - & 150k & 24.8k & 0 & 4.60k & - \\
        \cline{3-13}
            & & 8 & 112k & 66.5k & \REVA{1.69} & 7.07 & 7.88 & 19.9k & 6.26k & 16 & 578 & -  \\
        \cline{2-13}
            & \multirow{2}*{GAP8 + RVfplib}
            & 1 & 463k & 351k & \REVA{1.32} & - & - & 96.8k & 37 & 0 & 1 & - \\
        \cline{3-13}
            & & 8 & 67.8k & 45.5k & \REVA{1.49} & 6.83 & 7.95 & 11.6k & 3.88k & 12 & 4 & - \\
        \cline{2-13}
            & \multirow{2}*{PULP-OPEN}
            & 1 & 30.9k & 24.6k & \REVA{1.26} & - & - & 6.10k & 5 & 0 & 1 & 184 \\
        \cline{3-13}
            & & 8 & 4.66k & 3.34k & \REVA{1.39} & 6.63 & 7.88 & 766 & 283 & 198 & 3 & 80\\
        \hline
        \hline
        
        \multirow{6}*{GNB} 
            & \multirow{2}*{GAP8 + libgcc}
            & 1 & 86.4M & 67.4M & \REVA{1.28} & - & - & 15.9M & 3.38M & 0 & 16.1k & - \\
        \cline{3-13}
            & & 8 & 11.5M & 8.22M & \REVA{1.4} & 7.49 & 7.89 & 1.99M & 785k & 453 & 2.07k & - \\
        \cline{2-13}
            & \multirow{2}*{GAP8 + RVfplib}
            & 1 & 62.0M & 50.1M & \REVA{1.24} & - & - & 11M & 387k & 0 & 1 & - \\
        \cline{3-13}
            & & 8 & 8.09M & 6.09M & \REVA{1.33} & 7.64 & 7.96 & 1.37M & 299k & 507 & 62 & - \\
        \cline{2-13} 
            & \multirow{2}*{PULP-OPEN} 
            & 1 & 3.05M & 2.72M & \REVA{1.12} & - & - & 279k & 37.9k & 0 & 1 & 30.7k \\ 
        \cline{3-13}
            & & 8 & 463k & 345k & \REVA{1.34} & 6.56 & 7.91 & 34.7k & 16.8k & 1.49k & 62 & 44.1k \\
        \hline
        \hline
        
        \multirow{6}*{RF} 
            & \multirow{2}*{GAP8 + libgcc}
            & 1 & 1.01M & 695k & \REVA{1.45} & - & - & 344k & 39.9k & 0 & 1 & - \\
        \cline{3-13}
            & & 8 & 151k & 89.5k & \REVA{1.69} & 6.66 & 7.92 & 43.3k	& 11.4k & 420 & 60 & - \\
        \cline{2-13}
            & \multirow{2}*{GAP8 + RVfplib}
            & 1 & 742k & 629k & \REVA{1.18} & - & - & 78.8k & 18.5k & 0 & 1 & - \\
        \cline{3-13}
            & & 8 & 111k & 81.2k & \REVA{1.36} &	6.7 & 7.9 & 10.4k &	2.46k & 600 & 60	& - \\
        \cline{2-13}
            & \multirow{2}*{PULP-OPEN} 
            & 1 & 405k & 350k & \REVA{1.16} & - & - & 70.5k & 19.9k & 0 & 1 & 0 \\
        \cline{3-13}
            & & 8 & 59.4k & 44.1k & \REVA{1.35} & 6.82 & 7.81 & 9.16k & 1.32k & 1.08k & 60 & 0    \\
        \hline
        \hline
        
        \multirow{6}*{kNN} 
            & \multirow{2}*{GAP8 + libgcc}
            & 1 & 117M & 80.7M & \REVA{1.45} & - & - & 29.1M & 1.57M & 0 & 554k & - \\
        \cline{3-13}
            & & 8 & 15.4M & 10.1M & \REVA{1.52} &	7.59 & 7.94 & 3.64M 		& 	808k	&	1.58k	& 69.5k	& - \\
        \cline{2-13}
            & \multirow{2}*{GAP8 + RVfplib}
            & 1 & 61.6M & 46.5M & \REVA{1.32} & - & - & 13.3M & 635k & 0 & 15 & - \\
        \cline{3-13}
            & & 8 & 8.2M	& 5.84M	 & \REVA{1.4} & 	7.51 & 7.93 & 1.67M & 608k & 1.69k & 225 & -\\
        \cline{2-13}
            & \multirow{2}*{PULP-OPEN} 
            & 1 & 3.64M & 2.85M & \REVA{1.28} & - & - & 735k & 36.6k & 0 & 5 & 0\\
        \cline{3-13}
            & & 8 & 548k	& 377k	 & \REVA{1.45} & 6.65 & 7.59 & 91.4k 	& 7.09k &	858	& 225	& 253 \\
        \hline
        \hline
        
        \multirow{6}*{kMEANS} 
            & \multirow{2}*{GAP8 + libgcc}
            & 1 & 625k & 466k  & \REVA{1.34} & -  & -& 89.4k & 8.39M & 0 & 515 & - \\
        \cline{3-13}
            & & 8 & 83.6k & 59.3k & \REVA{1.41} & 7.47 & 8 & 12.7k & 3.66k & 9 & 98 & - \\
        \cline{2-13}
            & \multirow{2}*{GAP8 + RVfplib}
            & 1 & 395k & 315M & \REVA{1.25} & - & - & 45.4k & 525 & 0 & 1 & - \\
        \cline{3-13}
            & & 8 & 54.2k & 39.9k  & \REVA{1.36} &	7.29 & 8 & 6.83k &	2.62k & 10 & 1 & - \\
        \cline{2-13}
            & \multirow{2}*{PULP-OPEN}
            & 1 & 20.5k & 18.3k & \REVA{1.26} & - & - & 2.8k & 41 & 0 & 1 & 44\\
        \cline{3-13}
            & & 8 & 2.94k	& 2.17k	 & \REVA{1.35} & 6.98 & 8 & 353 	&	41	& 4	& 1 & 10 \\
        \hlineB{2}
    \end{tabular}
    }
    \label{tab:parallel_performance}
\end{table*}

To retrieve the highest predicted probability, GEMM-based kernels leverage the \(argmax\) sequential routine.
Thus, the theoretically achievable speedup decreases to 7.83\(\times\)-7.95\(\times\) depending on the deployed platform and FP emulation support.
The parallel algorithm design allows achieving speedups between 6.63\(\times\) and 7.07\(\times\) by switching the configuration.
By emulating FP computations on GAP8, I\$ misses do not scale linearly with the number of cores while increasing from almost zero to 5.72\% of the parallel execution time in LR with RVfplib support.
While other non-idealities are negligible, I\$ misses limit the speedup to \(7.07\times\) for libgcc and \(6.63\times\) for RVfplib when performing GEMM-based kernels on GAP8.
By leveraging the PULP-OPEN platform, the parallel computing time decreases to 4.20-4.66~kcycles making minor non-ideality sources affecting the performance.
Among the most significant, TCDM contentions represent 3.92-4.25\% of the PULP-OPEN 8-core execution time, highly bounding the speedup.
Moreover, I\$ misses increase when offloading the kernel computation onto CL. In particular, LR shows an I\$ misses rise from nearly zero to 6.08\% of the parallel runtime.
Regarding the FPU non-idealities, they explain up to 1.74\% of the parallel execution time, thus not limiting CL utilization.
However, despite the above-mentioned architectural factors, the optimized algorithm design allows reaching 6.63\(\times\)-7.05\(\times\) parallel speedup.

By emulating GNB FP computations on the GAP8 8-core CL, we improve the sequential execution by 7.49\(\times\) for libgcc FP support and 7.64\(\times\) for the custom RVfplib library. 
The architectural factor limiting the speedup on both emulation supports is related to I\$ misses since they slowly decrease moving to the parallel execution.
Performing the kernel on PULP-OPEN leads to not-negligible FPU non-idealities that double up compared to the sequential execution and
account for almost 10\% of the parallel runtime.
Concurrently, several architecture factors contribute to limiting CL compute efficiency, particularly I\$ misses do not scale linearly while covering 3.63\% of the parallel execution time. 
Therefore, leveraging the 8-core PULP-OPEN CL decreases GNB inference to 463~kcycles, thus reaching a 6.56x speedup compared to a single-core execution.

The most significant impact of architectural non-idealities involves a decrease in the CL performance efficiency when dispatching the RF kernel onto the 8-core engine.
By deploying libgcc to emulate FP comparison operations, the runtime reduces down 151~kcycles with a speedup of 6.66\(\times\).
Accordingly, RVfplib decreases the computing time from 742~kcycles to 111~kcycles enabling a 6.7\(\times\) performance improvement.
In addition to the sequential \(argmax\) routine limiting the gain to 7.9\(\times\) speedup, I\$ misses, and TCDM contentions bound the performance speedup accounting for 3\%-7\% of the parallel execution time.
Instead, PULP-OPEN achieves a 6.82\(\times\) computation time improvement compared to a single-core execution, presenting a theoretical speedup of 6.82\(\times\). 
The reduced kernel FLOP intensity (6.39\%) involves a low FPU usage, thus leading to zero FPU pipeline non-idealities.
\REVB{I\$ misses and TCDM contentions are the main architectural factors limiting the performance, impacting almost 4\% on the parallel computation time.}

Offloading kNN computations to the GAP8 8-core CL while deploying libgcc emulation support reduces the execution time from 117~Mcycles to 15.4~Mcycles, thus reaching a \(7.59\times\) speedup. 
Leveraging the optimized RVfplib library, kNN optimized parallel design improves the single-core running time by \(7.51\times\).
In both implementations, I\$ misses limits the CL compute power utilization since they scale sub-linearly with the number of cores while accounting for 5.24\%-7.41\% of the parallel execution time. 
By running the kernel on PULP-OPEN, we improve the runtime from 3.64~Mcycles to 548~kcycles leading to a 6.65\(\times\) speedup.
Due to PULP-OPEN reduced execution time, the sequential code weighs more on the computation and strictly limits the theoretical speedup to 7.59\(\times\) with 28~kcycles executed by a single-core.
Furthermore, architectural factors such as I\$ misses, TCDM contentions, and Ext-LD restrict the runtime reduction when offloading kNN computations to PULP-OPEN 8-core CL.

Considering the remaining MS-based algorithm, kMEANS features a 7.47\(\times\)-7.29\(\times\) runtime improvement compared to a sequential execution deploying libgcc and RVfplib on GAP8, respectively.
While the theoretical speedup attains almost 8\(\times\), architectural non-idealities limit the speedup when leveraging the 8-core CL.
I\$ misses account for a large portion of the parallel execution time, slowly decreasing in libgcc and growing from nearly zero to 4.83\% of the parallel computing time when deploying RVfplib emulation support. 
By switching to the PULP-OPEN platform, the FPU-native system decreases the 20.5~kcycles single-core execution time to 2.94~kcycles leveraging the 8-core CL.
Along with I\$ misses, several architectural factors such as TCDM contentions and Ext-LD contributes to bounding the speedup improvement to 6.98\(\times\).

\REVA{Adopting optimized parallel designs of Non-Neural ML kernels on top of PULP processors also offers several benefits for latency and energy efficiency, which are crucial in the TinyML domain. By fully leveraging the 8-core CL compute power, we enable performing the kernels with an excellent latency and energy decrease ranging from 85\% to 87\%. Executing Parallel k-Means and GEMM-based algorithms on the PULP-OPEN platform requires only 7.35-11~\(\mu\)s latency and 0.36-0.55~\(\mu\)J per inference. While RF demands 149~\(\mu\)s and 7.34~\(\mu\)J, dispatching NB and kNN onto the 8-core CL reduces the latency and energy usage to 1.2-1.4~\(m\)s and 57-67~\(\mu\)J.}

\begin{figure}[t]
    \centering
    \captionsetup{justification=centering}
    \includegraphics[width=0.4\linewidth]{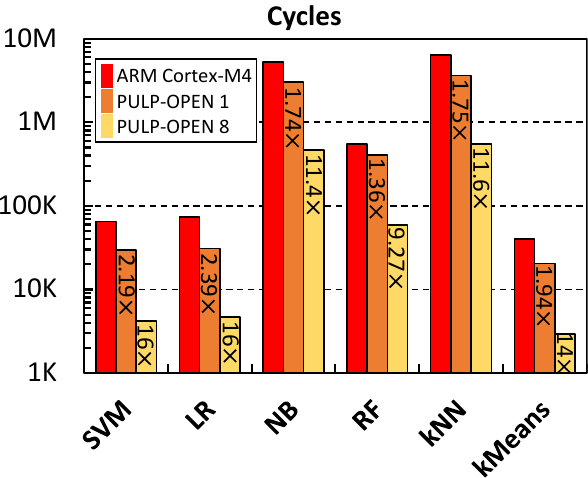}
    \caption{\REVB{ARM Cortex-M4 vs. PULP-OPEN comparison.}}
    \label{fig:ARM_COMPARISON}
\end{figure}
\subsection{Comparison with Cortex-M4}
\label{ARM}
This section compares the execution time of the Non-Neural ML kernels between PULP-OPEN and the ARM Cortex-M4\footnote{\url{https://developer.arm.com/Processors/Cortex-M4}} architecture.
\REVB{This comparison focuses on single-core sequential execution because the techniques proposed for code parallelization require minimal runtime support and are, to a large degree, orthogonal to the ISA and the core micro-architecture.}
\REVB{We used for comparison an STM32F4\footnote{\url{https://www.st.com/en/microcontrollers-microprocessors/stm32f4-series.html}} MCU since it belongs to a widespread, commercially successful ultra-low-power MCU family. The STM32F4 features the Adaptive Real-Time (ART) memory accelerator to speed up instructions fetch along with DSP and FPU instructions support.}
To perform the experimental evaluation, we optimized the Non-Neural ML algorithms for the Cortex-M4 target using CMSIS-DSP routines and custom-coded functions not provided in the library.
In particular, we leveraged CMSIS-DSP GNB and linear SVM implementations while the LR design for  Cortex-M4 uses the optimized dot product included in the library.
CMSIS-DSP Euclidean distance routine embeds the square root calculation.
Thus, we improved the distance metric by removing such a multi-cycle operation in MS-based algorithms.
Since there is no CMSIS-DSP support for RF, we coded the kernel for the STM32F4 target \REVB{using the same optimization strategies we devised for the sequential implementation on PULP.}

\REVB{
Figure~\ref{fig:ARM_COMPARISON} reports the cycles required for the sequential execution of the ML benchmarks on Cortex-M4 and PULP-OPEN. 
The figure also reports the results executing on the 8-core CL as a further reference.
We report the achieved speedup w.r.t. the Cortex-M4 on top of the bars.
Focusing on the sequential execution, PULP-OPEN achieves speedups ranging from 1.36\(\times\) to 2.39\(\times\) compared to Cortex-M4.
While RF execution on PULP-OPEN achieves a 1.36\(\times\) execution time decrease, GEMM-based kernels reach up to a 2.39\(\times\) runtime improvement.
Along with GNB, MS-based algorithms attain an intermediate improvement result with a 1.74\(\times\)-1.94\(\times\) speedup.

Both architectures execute kernels optimized explicitly for their ISA, and execution time is expressed in cycles (i.e., it is independent of frequency). This gap is due to three main factors: single-cycle load operations, hardware loop support, and fused multiply-and-add FP operations. Load operations are executed in a single cycle when programmers adopt techniques to reduce data dependencies inside the loop body (e.g., loop unrolling). Adopting hardware loops saves one register, removes the overhead of updating the loop counter, and avoids pipeline stalls when the branch is taken. Finally, multiply-and-accumulate operations require two cycles on PULP-OPEN, but they are pipelined so that the throughput is close to 1 op/cycle when the compiler avoids data dependencies on the output register.
}
\REVA{
}

\REVA{
}

%% file: 06-conclusion.tex
\section{Conclusion}
This paper presents the parallel design of six relevant Non-Neural ML algorithms to fit ML computational constraints into edge-based PULP MCUs.
We developed the algorithm design targetting efficient execution on GAP8, a commercial chip, and PULP-OPEN, a research platform running on an FPGA emulator. 
We determined efficient memory access patterns and parallelization schemes achieving peak performance by optimizing the runtime through a fine-grained analysis and extensive optimization.
Since IoT-class MCUs often limit the HW resources to benefit energy efficiency, we leveraged two alternative FP emulation libraries to perform FP computations on the FPU-less GAP8.

By comparing the Non-Neural ML kernels execution time on a single-core GAP8 configuration, we show that the target-optimized RVfplib library achieves an average 1.61\(\times\) speedup compared to the standard libgcc emulation support. 
Instead, leveraging the FPU-native support on a single-core PULP-OPEN allows up to 32.09\(\times\) speedup compared to libgcc emulation.
We also examined the parallel performance on the adopted PULP platforms, comparing the single-core execution time with the 8-core CL runtime.
The parallel design enables near-ideal speedups ranging from 6.56\(\times\) to 7.64\(\times\), considering the two PULP platforms and GAP8 FP emulation supports.
We support the discussion with a comprehensive runtime analysis providing core- and SoC-level architectural factors limiting the speedup in each platform configuration and algorithm.
Lastly, we present a comparison between PULP-OPEN and ARM Cortex-M4.
By leveraging PULP-OPEN in a single-core configuration, we achieve 1.36\(\times\)-2.39\(\times\) speedup compared to Cortex-M4 deployment.
At the same time, using the 8-core CL of PULP-OPEN reduces the runtime drastically, leading to a 9.27\(\times\)-15.85\(\times\) performance improvement.

Future work will include the design of an automatic tool to deploy Non-Neural ML algorithms on PULP-based MCUs targetting optimal tiling and double-buffering operations to achieve peak performance.
Furthermore, we will expand the developed parallel library by integrating further Non-Neural ML kernels and supporting new emerging PULP architectures.

%% file: main.bbl
\begin{thebibliography}{10}

\bibitem{evans2011IoT}
D.~Evans.
\newblock {The Internet of Things: How the Next Evolution of the Internet Is
  Changing Everything}.
\newblock Technical report, Cisco, 2011.

\bibitem{dogan2011autonomous}
Ürün Dogan, Johann Edelbrunner, and Ioannis Iossifidis.
\newblock Autonomous driving: A comparison of machine learning techniques by
  means of the prediction of lane change behavior.
\newblock In {\em 2011 IEEE International Conference on Robotics and
  Biomimetics}, pages 1837--1843. IEEE, 2011.

\bibitem{tabanelli2021trimming}
Enrico Tabanelli, Davide Brunelli, Andrea Acquaviva, and Luca Benini.
\newblock {Trimming Feature Extraction and Inference for MCU-based Edge NILM: a
  Systematic Approach}.
\newblock {\em IEEE Transactions on Industrial Informatics}, 18(2):943--952,
  2022.

\bibitem{ml2017ku}
Pradeep Kumar, Pradeep Kumar, and Arvind Tiwari.
\newblock {\em {Ubiquitous Machine Learning and Its Applications}}.
\newblock IGI Global, USA, 1st edition, 2017.

\bibitem{cisco2011IoT}
Cisco.
\newblock {Global Cloud Index: Forecast and Methodology, 2016–2021}.
\newblock Technical report, Cisco, 2016.

\bibitem{barbera2013offload}
Marco~V Barbera, Sokol Kosta, Alessandro Mei, and Julinda Stefa.
\newblock {To offload or not to offload? the bandwidth and energy costs of
  mobile cloud computing}.
\newblock In {\em 2013 Proceedings IEEE Infocom}, pages 1285--1293. IEEE, 2013.

\bibitem{sun2014data}
Yunchuan Sun, Junsheng Zhang, Yongping Xiong, and Guangyu Zhu.
\newblock {Data security and privacy in cloud computing}.
\newblock {\em International Journal of Distributed Sensor Networks},
  10(7):190903, 2014.

\bibitem{sanchez2020tinyml}
Ramon Sanchez-Iborra and Antonio~F Skarmeta.
\newblock {TinyML-Enabled Frugal Smart Objects: Challenges and Opportunities}.
\newblock {\em IEEE Circuits and Systems Magazine}, 20(3):4--18, 2020.

\bibitem{banbury2020benchmarking}
Colby~R Banbury, Vijay~Janapa Reddi, Max Lam, William Fu, Amin Fazel, Jeremy
  Holleman, Xinyuan Huang, Robert Hurtado, David Kanter, Anton Lokhmotov,
  et~al.
\newblock {Benchmarking TinyML systems: Challenges and direction}.
\newblock {\em arXiv preprint arXiv:2003.04821}, 2020.

\bibitem{tinyML}
{TinyML foundation}.
\newblock {TinyML reasearch community}.
\newblock \url{https://www.tinyml.org/}, last accessed on {2022-10-15}.

\bibitem{yu2017survey}
Wei Yu, Fan Liang, Xiaofei He, William~Grant Hatcher, Chao Lu, Jie Lin, and
  Xinyu Yang.
\newblock {A survey on the edge computing for the Internet of Things}.
\newblock {\em IEEE Access}, 6:6900--6919, 2017.

\bibitem{he2016deep}
Kaiming He, Xiangyu Zhang, Shaoqing Ren, and Jian Sun.
\newblock {Deep residual learning for image recognition}.
\newblock In {\em Proceedings of the IEEE conference on computer vision and
  pattern recognition}, pages 770--778, 2016.

\bibitem{tan2019efficientnet}
Mingxing Tan and Quoc Le.
\newblock {EfficientNet: Rethinking Model Scaling for Convolutional Neural
  Networks}.
\newblock In {\em International Conference on Machine Learning}, pages
  6105--6114. PMLR, 2019.

\bibitem{sandler2018mobilenetv2}
Mark Sandler, Andrew Howard, Menglong Zhu, Andrey Zhmoginov, and Liang-Chieh
  Chen.
\newblock {Mobilenetv2: Inverted residuals and linear bottlenecks}.
\newblock In {\em Proceedings of the IEEE conference on computer vision and
  pattern recognition}, pages 4510--4520, 2018.

\bibitem{gapprocessors}
{Greenwaves Technologies}.
\newblock {GAP Processors}.
\newblock \url{https://greenwaves-technologies.com/gap8_gap9/}, last accessed
  on {2022-10-15}.

\bibitem{spresensesony}
{Sony}.
\newblock {Spresense development board}.
\newblock \url{https://developer.sony.com/develop/spresense/}, last accessed on
  {2022-10-15}.

\bibitem{mittal2015survey}
Sparsh Mittal.
\newblock {A survey of architectural techniques for near-threshold computing}.
\newblock {\em ACM Journal on Emerging Technologies in Computing Systems
  (JETC)}, 12(4):1--26, 2015.

\bibitem{flamand2018gap8}
E.~Flamand, D.~Rossi, F.~Conti, I.~Loi, A.~Pullini, F.~Rotenberg, , and
  L.~Benini.
\newblock {GAP-8: A RISC-V SoC for AI at the Edge of the IoT }.
\newblock In {\em International Conference on Application-specific Systems,
  Architectures and Processors (ASAP)}, pages 1--4. IEEE, 2018.

\bibitem{rossi2021vega}
Davide Rossi, Francesco Conti, Manuel Eggiman, Stefan Mach, Alfio~Di Mauro,
  Marco Guermandi, Giuseppe Tagliavini, Antonio Pullini, Igor Loi, Jie Chen,
  Eric Flamand, and Luca Benini.
\newblock {4.4 A 1.3TOPS/W @ 32GOPS Fully Integrated 10-Core SoC for IoT
  End-Nodes with 1.7\(\mu\)W Cognitive Wake-Up From MRAM-Based State-Retentive
  Sleep Mode}.
\newblock In {\em 2021 IEEE International Solid- State Circuits Conference
  (ISSCC)}, volume~64, pages 60--62, 2021.

\bibitem{gottscho2017low}
Mark Gottscho, Irina Alam, Clayton Schoeny, Lara Dolecek, and Puneet Gupta.
\newblock {Low-cost memory fault tolerance for IoT devices}.
\newblock {\em ACM Transactions on Embedded Computing Systems (TECS)},
  16(5s):1--25, 2017.

\bibitem{doris2013profile}
Doris Chen and Deshanand Singh.
\newblock {Profile-Guided Floating- to Fixed-Point Conversion for Hybrid
  FPGA-Processor Applications}.
\newblock {\em ACM Transactions on Architecture and Code Optimization}, 9(4),
  2013.

\bibitem{menard2006floating}
Daniel Menard, Daniel Chillet, and Olivier Sentieys.
\newblock {Floating-to-fixed-point conversion for digital signal processors}.
\newblock {\em EURASIP Journal on Advances in Signal Processing}, 2006:1--19,
  2006.

\bibitem{christensen2006fixed}
Michael Christensen and Fred~J Taylor.
\newblock {Fixed-point-IIR-filter challenges}.
\newblock {\em EDN Netw}, 51(23):111--122, 2006.

\bibitem{menard2008accuracy}
Daniel Menard, Romain Serizel, Romuald Rocher, and Olivier Sentieys.
\newblock Accuracy constraint determination in fixed-point system design.
\newblock {\em EURASIP Journal on Embedded Systems}, 2008:1--12, 2008.

\bibitem{chang2008fft}
Wei-Hsin Chang and Truong~Q. Nguyen.
\newblock {On the Fixed-Point Accuracy Analysis of FFT Algorithms}.
\newblock {\em IEEE Transactions on Signal Processing}, 56(10):4673--4682,
  2008.

\bibitem{perotti2021rvfplib}
Matteo Perotti, Giuseppe Tagliavini, Stefan Mach, Luca Bertaccini, and Luca
  Benini.
\newblock {RVfplib: A Fast and Compact Open-Source Floating-Point Emulation
  Library for Tiny RISC-V Processors}.
\newblock In {\em International Conference on Embedded Computer Systems}, pages
  16--32. Springer, 2022.

\bibitem{capra2020hwswnn}
Maurizio Capra, Beatrice Bussolino, Alberto Marchisio, Guido Masera, Maurizio
  Martina, and Muhammad Shafique.
\newblock {Hardware and Software Optimizations for Accelerating Deep Neural
  Networks: Survey of Current Trends, Challenges, and the Road Ahead}.
\newblock {\em IEEE Access}, 8:225134--225180, 2020.

\bibitem{greeshma2019fashion}
KV~Greeshma and K~Sreekumar.
\newblock {Fashion-MNIST classification based on HOG feature descriptor using
  SVM}.
\newblock {\em International Journal of Innovative Technology and Exploring
  Engineering (IJITEE)}, 8(5):960--962, 2019.

\bibitem{lecun1998gradient}
Yann LeCun, L{\'e}on Bottou, Yoshua Bengio, and Patrick Haffner.
\newblock {Gradient-based learning applied to document recognition}.
\newblock {\em Proceedings of the IEEE}, 86(11):2278--2324, 1998.

\bibitem{lai2018cmsis}
Liangzhen Lai, Naveen Suda, and Vikas Chandra.
\newblock {CMSIS-NN: Efficient Neural Network Kernels for Arm Cortex-M CPUs}.
\newblock {\em arXiv preprint arXiv:1801.06601}, 2018.

\bibitem{x-cube-ai}
{\relax STMicroelectronics}.
\newblock {X-Cube-AI: AI Expansion Pack for STM32CubeMX}.
\newblock \url{https://www.st.com/en/embedded-software/x-cube-ai.html}, last
  accessed on 2022-10-15.

\bibitem{yazici2018edge}
Mahmut~Taha Yazici, Shadi Basurra, and Mohamed~Medhat Gaber.
\newblock {Edge Machine Learning: Enabling Smart Internet of Things
  Applications}.
\newblock {\em Big data and cognitive computing. MDPI}, 2(3):26, 2018.

\bibitem{bekaroo2016pi}
Girish Bekaroo and Aditya Santokhee.
\newblock {Power consumption of the Raspberry Pi: A comparative analysis}.
\newblock In {\em 2016 IEEE International Conference on Emerging Technologies
  and Innovative Business Practices for the Transformation of Societies
  (EmergiTech)}, pages 361--366, 2016.

\bibitem{sakr2020machine}
Fouad Sakr, Francesco Bellotti, Riccardo Berta, and Alessandro De~Gloria.
\newblock {Machine Learning on Mainstream Microcontrollers}.
\newblock {\em Sensors. MDPI}, 20(9):2638, 2020.

\bibitem{microml}
{\relax Eloquent Arduino blog}.
\newblock {MicroML}.
\newblock \url{https://github.com/eloquentarduino/micromlgen}, last accessed on
  {2022-10-15}.

\bibitem{emlearn}
Jon Nordby.
\newblock {Emlearn: Machine Learning inference engine for Microcontrollers and
  Embedded Devices}.
\newblock \url{https://github.com/emlearn/emlearn}, last accessed on
  {2022-10-15}.

\bibitem{almansoor2020parallel}
Mohamed Almansoor, Mohamed Alaradi, and Abdulla Alqaddoumi.
\newblock {Parallel Programming for Classification Algorithms Using Logistic
  Regression and Artificial Neural Networks: Framework and Applications}.
\newblock In {\em 2020 International Conference on Data Analytics for Business
  and Industry: Way Towards a Sustainable Economy (ICDABI)}, pages 1--6. IEEE,
  2020.

\bibitem{senagi2022parallel}
Kennedy Senagi and Nicolas Jouandeau.
\newblock {Parallel construction of Random Forest on GPU}.
\newblock {\em The Journal of Supercomputing. Springer}, pages 1--21, 2022.

\bibitem{liu2019parallel}
Peng Liu, Hui-han Zhao, Jia-yu Teng, Yan-yan Yang, Ya-feng Liu, and Zong-wei
  Zhu.
\newblock {Parallel naive Bayes algorithm for large-scale Chinese text
  classification based on Spark}.
\newblock {\em Journal of Central South University. Springer}, 26(1):1--12,
  2019.

\bibitem{you2014mic}
Yang You, Shuaiwen~Leon Song, Haohuan Fu, Andres Marquez, Maryam~Mehri Dehnavi,
  Kevin Barker, Kirk~W Cameron, Amanda~Peters Randles, and Guangwen Yang.
\newblock {Mic-svm: Designing a highly efficient support vector machine for
  advanced modern multi-core and many-core architectures}.
\newblock In {\em 2014 IEEE 28th International Parallel and Distributed
  Processing Symposium}, pages 809--818. IEEE, 2014.

\bibitem{zhu2019high}
Huming Zhu, Pei Li, Peng Zhang, and Zheng Luo.
\newblock {A High Performance Parallel Ranking SVM with OpenCL on Multi-core
  and Many-core Platforms}.
\newblock {\em International Journal of Grid and High Performance Computing
  (IJGHPC). IGI Global}, 11(1):17--28, 2019.

\bibitem{ma2019stochastic}
Yujing Ma, Florin Rusu, and Martin Torres.
\newblock {Stochastic gradient descent on modern hardware: Multi-core CPU or
  GPU? Synchronous or asynchronous?}
\newblock In {\em 2019 IEEE International Parallel and Distributed Processing
  Symposium (IPDPS)}, pages 1063--1072. IEEE, 2019.

\bibitem{gupta2017protonn}
Chirag Gupta, Arun~Sai Suggala, Ankit Goyal, Harsha~Vardhan Simhadri, Bhargavi
  Paranjape, Ashish Kumar, Saurabh Goyal, Raghavendra Udupa, Manik Varma, and
  Prateek Jain.
\newblock {ProtoNN: Compressed and Accurate kNN for Resource-scarce Devices}.
\newblock In {\em International Conference on Machine Learning}, pages
  1331--1340. PMLR, 2017.

\bibitem{kumar2017resource}
Ashish Kumar, Saurabh Goyal, and Manik Varma.
\newblock {Resource-efficient Machine Learning in 2 KB RAM for the Internet of
  Things}.
\newblock In {\em International Conference on Machine Learning}, pages
  1935--1944. PMLR, 2017.

\bibitem{gopinath2019kbsized}
Sridhar Gopinath, Nikhil Ghanathe, Vivek Seshadri, and Rahul Sharma.
\newblock {Compiling KB-Sized Machine Learning Models to Tiny IoT Devices}.
\newblock In {\em Proceedings of the 40th ACM SIGPLAN Conference on Programming
  Language Design and Implementation}, page 79–95. ACM, 2019.

\bibitem{mahajan2016tabla}
Divya Mahajan, Jongse Park, Emmanuel Amaro, Hardik Sharma, Amir Yazdanbakhsh,
  Joon~Kyung Kim, and Hadi Esmaeilzadeh.
\newblock {Tabla: A unified template-based framework for accelerating
  statistical machine learning}.
\newblock In {\em 2016 IEEE International Symposium on High Performance
  Computer Architecture (HPCA)}, pages 14--26. IEEE, 2016.

\bibitem{MAHDAVINEJAD2018161}
Mohammad~Saeid Mahdavinejad, Mohammadreza Rezvan, Mohammadamin Barekatain,
  Peyman Adibi, Payam Barnaghi, and Amit~P. Sheth.
\newblock {Machine learning for internet of things data analysis: a survey}.
\newblock {\em Digital Communications and Networks}, 4(3):161--175, 2018.

\bibitem{merenda2020edge}
Massimo Merenda, Carlo Porcaro, and Demetrio Iero.
\newblock {Edge machine learning for AI-enabled IoT devices: A review}.
\newblock {\em Sensors}, 20(9):2533, 2020.

\bibitem{ahmad2017trees}
Muhammad~Waseem Ahmad, Monjur Mourshed, and Yacine Rezgui.
\newblock {Trees vs Neurons: Comparison between random forest and ANN for
  high-resolution prediction of building energy consumption}.
\newblock {\em Energy and buildings}, 147:77--89, 2017.

\bibitem{banbury2021mlperf}
Colby Banbury, Vijay~Janapa Reddi, Peter Torelli, Jeremy Holleman, Nat
  Jeffries, Csaba Kiraly, Pietro Montino, David Kanter, Sebastian Ahmed, Danilo
  Pau, et~al.
\newblock {MLPerf Tiny Benchmark}.
\newblock {\em arXiv preprint arXiv:2106.07597}, 2021.

\bibitem{huh2021metric}
Jaesung Huh, Minjae Lee, Heesoo Heo, Seongkyu Mun, and Joon~Son Chung.
\newblock {Metric Learning for Keyword Spotting}.
\newblock In {\em 2021 IEEE Spoken Language Technology Workshop (SLT)}, pages
  133--140. IEEE, 2021.

\bibitem{shor2022universal}
Joel Shor, Aren Jansen, Wei Han, Daniel Park, and Yu~Zhang.
\newblock {Universal paralinguistic speech representations using
  self-supervised conformers}.
\newblock In {\em ICASSP 2022-2022 IEEE International Conference on Acoustics,
  Speech and Signal Processing (ICASSP)}, pages 3169--3173. IEEE, 2022.

\bibitem{liu2019fusing}
Xueliang Liu, Rongjie Zhang, Zhijun Meng, Richang Hong, and Guangcan Liu.
\newblock {On fusing the latent deep CNN feature for image classification}.
\newblock {\em World Wide Web}, 22(2):423--436, 2019.

\bibitem{durkota2020neuron}
Karel Durkota, Michal Linda, M~Ludvik, and Jan Tozicka.
\newblock {Neuron-net: Siamese network for anomaly detection}.
\newblock Technical report, DCASE2020 Challenge, Tech. Rep, 2020.

\bibitem{zhao2021spatiotemporal}
Minglu Zhao, Hiroyuki Takizawa, and Tomoya Soma.
\newblock {Spatiotemporal Anomaly Detection for Large-Scale Sensor Data}.
\newblock In {\em 2021 12th International Symposium on Parallel Architectures,
  Algorithms and Programming (PAAP)}, pages 162--168. IEEE, 2021.

\bibitem{dagun1998openmp}
L.~Dagum and R.~Menon.
\newblock {OpenMP: an industry standard API for shared-memory programming}.
\newblock {\em IEEE Computational Science and Engineering}, 5(1):46--55, 1998.

\bibitem{mach2021fpnew}
S.~Mach, F.~Schuiki, F.~Zaruba, and L.~Benini.
\newblock {FPnew: An Open-Source Multiformat Floating-Point Unit Architecture
  for Energy-Proportional Transprecision Computing}.
\newblock {\em IEEE Transactions on Very Large Scale Integration (VLSI)
  Systems}, 29(4):774--787, 2021.

\bibitem{nntile2020zhuo}
Xiaoyan Zhuo, Iman Nandi, Taha Azzaoui, and Seung~Woo Son.
\newblock {A Neural Network-Based Optimal Tile Size Selection Model for
  Embedded Vision Applications}.
\newblock In {\em 2020 IEEE 22nd International Conference on High Performance
  Computing and Communications; IEEE 18th International Conference on Smart
  City; IEEE 6th International Conference on Data Science and Systems
  (HPCC/SmartCity/DSS)}, pages 607--612, 2020.

\bibitem{burrello2021dory}
Alessio Burrello, Angelo Garofalo, Nazareno Bruschi, Giuseppe Tagliavini,
  Davide Rossi, and Francesco Conti.
\newblock {DORY: Automatic End-to-End Deployment of Real-World DNNs on Low-Cost
  IoT MCUs}.
\newblock {\em IEEE Transactions on Computers}, 2021.

\bibitem{chandra2001parallel}
Rohit Chandra, Leo Dagum, David Kohr, Ramesh Menon, Dror Maydan, and Jeff
  McDonald.
\newblock {\em {Parallel programming in OpenMP}}.
\newblock Morgan kaufmann, 2001.

\bibitem{tagliavini2018unleashing}
Giuseppe Tagliavini, Daniele Cesarini, and Andrea Marongiu.
\newblock {Unleashing fine-grained parallelism on embedded many-core
  accelerators with lightweight OpenMP tasking}.
\newblock {\em IEEE Transactions on Parallel and Distributed Systems},
  29(9):2150--2163, 2018.

\bibitem{munera2020towards}
Adrian Munera, Sara Royuela, and Eduardo Qui{\~n}ones.
\newblock {Towards a qualifiable OpenMP framework for embedded systems}.
\newblock In {\em 2020 Design, Automation \& Test in Europe Conference \&
  Exhibition (DATE)}, pages 903--908. IEEE, 2020.

\bibitem{chapman2009implementing}
Barbara Chapman, Lei Huang, Eric Biscondi, Eric Stotzer, Ashish Shrivastava,
  and Alan Gatherer.
\newblock {Implementing OpenMP on a high performance embedded multicore MPSoC}.
\newblock In {\em 2009 IEEE International Symposium on Parallel \& Distributed
  Processing}, pages 1--8. IEEE, 2009.

\bibitem{agathos2013deploying}
Spiros~N Agathos, Vassilios~V Dimakopoulos, Aggelos Mourelis, and Alexandros
  Papadogiannakis.
\newblock {Deploying OpenMP on an embedded multicore accelerator}.
\newblock In {\em 2013 International Conference on Embedded Computer Systems:
  Architectures, Modeling, and Simulation (SAMOS)}, pages 180--187. IEEE, 2013.

\bibitem{patel2015survey}
Sumit Patel, MB~Potdar, and Bhadreshsinh Gohil.
\newblock {A survey on image processing techniques with OpenMP}.
\newblock {\em International Journal of Engineering Development and Research},
  3(4):837--839, 2015.

\bibitem{padilla2020detection}
Dionis~A Padilla, Ramon Alfredo~I Pajes, and Jerome~T De~Guzman.
\newblock {Detection of Corn Leaf Diseases Using Convolutional Neural Network
  With OpenMP Implementation}.
\newblock In {\em 2020 IEEE 12th International Conference on Humanoid,
  Nanotechnology, Information Technology, Communication and Control,
  Environment, and Management (HNICEM)}, pages 1--6. IEEE, 2020.

\bibitem{huang2012parallelizing}
Lei Huang, Eric Stotzer, Hangjun Yi, Barbara Chapman, and Sunita
  Chandrasekaran.
\newblock {Parallelizing ultrasound image processing using OpenMP on multicore
  embedded systems}.
\newblock In {\em 2012 IEEE Global High Tech Congress on Electronics}, pages
  131--138. IEEE, 2012.

\bibitem{furlinger2006analyzing}
Karl F{\"u}rlinger and Michael Gerndt.
\newblock {Analyzing overheads and scalability characteristics of OpenMP
  applications}.
\newblock In {\em International Conference on High Performance Computing for
  Computational Science}, pages 39--51. Springer, 2006.

\bibitem{darema2001spmd}
Frederica Darema.
\newblock {The SPMD model: Past, Present and Future}.
\newblock In {\em European Parallel Virtual Machine/Message Passing Interface
  Users’ Group Meeting}, pages 1--1. Springer, 2001.

\bibitem{montagna2021streamlining}
Fabio Montagna, Giuseppe Tagliavini, Davide Rossi, Angelo Garofalo, and Luca
  Benini.
\newblock {Streamlining the OpenMP Programming Model on Ultra-Low-Power
  Multi-core MCUs}.
\newblock In {\em International Conference on Architecture of Computing
  Systems}, pages 167--182. Springer, 2021.

\bibitem{cramer2002lr}
J.S. Cramer.
\newblock {The Origins of Logistic Regression}.
\newblock In {\em Tinbergen Institute Discussion}. Tinbergen Institute, 2002.

\bibitem{ioannou2018wsn}
Christiana Ioannou and Vasos Vassiliou.
\newblock {An Intrusion Detection System for Constrained WSN and IoT Nodes
  Based on Binary Logistic Regression}.
\newblock In {\em Proceedings of the 21st ACM International Conference on
  Modeling, Analysis and Simulation of Wireless and Mobile Systems}, page
  259–263. ACM, 2018.

\bibitem{HASAN2019100059}
Mahmudul Hasan, Md.~Milon Islam, Md~Ishrak~Islam Zarif, and M.M.A. Hashem.
\newblock {Attack and anomaly detection in IoT sensors in IoT sites using
  machine learning approaches}.
\newblock {\em Internet of Things}, 7:100059, 2019.

\bibitem{cortes1995svm}
V.~Vapnik C.~Cortes.
\newblock {Support-Vector Networks}.
\newblock {\em Machine learning}, 20(1):273--297, 1995.

\bibitem{svm2007yi}
Yi-Hung Liu and Yen-Ting Chen.
\newblock {Face Recognition Using Total Margin-Based Adaptive Fuzzy Support
  Vector Machines}.
\newblock {\em IEEE Transactions on Neural Networks}, 18(1):178--192, 2007.

\bibitem{svm2020sid}
T.~Siddharth, Pranjali Gajbhiye, Rajesh~Kumar Tripathy, and Ram~Bilas Pachori.
\newblock {EEG-Based Detection of Focal Seizure Area Using FBSE-EWT Rhythm and
  SAE-SVM Network}.
\newblock {\em IEEE Sensors Journal}, 20(19):11421--11428, 2020.

\bibitem{friedman1997nb}
Friedman Nir, Geiger Dan, and Goldszmidt Moises.
\newblock {Bayesian Network Classifiers}.
\newblock {\em Machine learning}, 29(7):131--163, 1997.

\bibitem{gnb2020wu}
Di~Wu, Zhongkai Jiang, Xiaofeng Xie, Xuetao Wei, Weiren Yu, and Renfa Li.
\newblock {LSTM Learning With Bayesian and Gaussian Processing for Anomaly
  Detection in Industrial IoT}.
\newblock {\em IEEE Transactions on Industrial Informatics}, 16(8):5244--5253,
  2020.

\bibitem{veh2021ku}
Nikhil Kumar, Debopam Acharya, and Divya Lohani.
\newblock {An IoT-Based Vehicle Accident Detection and Classification System
  Using Sensor Fusion}.
\newblock {\em IEEE Internet of Things Journal}, 8(2):869--880, 2021.

\bibitem{cover1967knn}
T.~Cover and P.~Hart.
\newblock {Nearest neighbor pattern classification}.
\newblock {\em IEEE Transactions on Information Theory}, 13(1):21--27, 1967.

\bibitem{gest2021wo}
W.~K. Wong, Filbert~H. Juwono, and Brendan Teng~Thiam Khoo.
\newblock {Multi-Features Capacitive Hand Gesture Recognition Sensor: A Machine
  Learning Approach}.
\newblock {\em IEEE Sensors Journal}, 21(6):8441--8450, 2021.

\bibitem{bone2019ran}
Ranjitha M~M, Taranath~N L, Arpitha~C N, and C.K. Subbaraya.
\newblock {Bone Cancer Detection Using K-Means Segmentation and Knn
  Classification}.
\newblock In {\em 2019 1st International Conference on Advances in Information
  Technology (ICAIT)}, pages 76--80, 2019.

\bibitem{macqueen1997kmeans}
J.~MacQueen.
\newblock {Some methods for classification and analysis of multivariate
  observations}.
\newblock In {\em Proceedings of the Fifth Berkeley Symposium on Mathematical
  Statistics and Probability: Weather modification}, pages 281--296. University
  of California, 1967.

\bibitem{data2017wu}
Wenbin Wu and Mugen Peng.
\newblock {A Data Mining Approach Combining $K$ -Means Clustering With Bagging
  Neural Network for Short-Term Wind Power Forecasting}.
\newblock {\em IEEE Internet of Things Journal}, 4(4):979--986, 2017.

\bibitem{pattern2013peng}
Xiaosheng Peng, Chengke Zhou, Donald~M. Hepburn, Martin~D. Judd, and W.~H.
  Siew.
\newblock {Application of K-Means method to pattern recognition in on-line
  cable partial discharge monitoring}.
\newblock {\em IEEE Transactions on Dielectrics and Electrical Insulation},
  20(3):754--761, 2013.

\bibitem{breiman2001rf}
L.~Breiman.
\newblock {Random Forests}.
\newblock {\em Machine learning}, 45(1):5--32, 2001.

\bibitem{tabanelli2020feature}
Enrico Tabanelli, Davide Brunelli, and Luca Benini.
\newblock {A Feature Reduction Strategy For Enabling Lightweight Non-Intrusive
  Load Monitoring On Edge Devices}.
\newblock In {\em 2020 IEEE 29th International Symposium on Industrial
  Electronics (ISIE)}, pages 805--810. IEEE, 2020.

\bibitem{ad2020lin}
Tzu-Hsuan Lin and Jehn-Ruey Jiang.
\newblock {Anomaly Detection with Autoencoder and Random Forest}.
\newblock In {\em 2020 International Computer Symposium (ICS)}, pages 96--99,
  2020.

\end{thebibliography}
